\documentclass[aps,superscriptaddress,twocolumn,pre,longbibliography,floatfix]{revtex4-1}

\usepackage{amsmath,amsthm,amsfonts,amssymb,bm,graphicx,color,mathpazo,times, braket, ulem}
\usepackage[colorlinks={true}, citecolor={blue}, filecolor={red}, linkcolor={red}, urlcolor={blue}]{hyperref}
\usepackage[caption=false]{subfig}
\usepackage{soul}
\usepackage{fix-cm} % Ensures all font shapes and sizes are available
\begin{document}
%------------------------------------------------------------------------%
 \title{Mixing thermal coherent states for precision and range enhancement in quantum thermometry}
	\author{Asghar Ullah}
 \email{aullah21@ku.edu.tr}
\affiliation{Department of Physics, Ko\c{c} University, 34450 Sar\i yer, Istanbul, T\"urkiye}
	\author{M. Tahir Naseem} 
 \email{mnaseem16@ku.edu.tr}
\affiliation{Faculty of Engineering Science, Ghulam Ishaq Khan Institute of Engineering Sciences and Technology, \\
Topi 23640, Khyber Pakhtunkhwa, Pakistan}
\author{\"Ozg\"ur E. M\"ustecapl\i o\u glu}	
	\email{omustecap@ku.edu.tr}
	\affiliation{Department of Physics, Ko\c{c} University, 34450 Sar\i yer, Istanbul, T\"urkiye}
 \affiliation{Faculty of Engineering and Natural Sciences, Sabancı University, Tuzla, Istanbul 34956, T\"urkiye}
 \affiliation{T\"UBITAK Research Institute for Fundamental Sciences, 41470 Gebze, T\"urkiye}
 \date{\today}
%---------------------------------------------------------------------------------
\begin{abstract}
The unavoidable interaction between thermal environments and quantum systems typically leads to the degradation of quantum coherence, which can be fought against by reservoir engineering. We propose the realization of a special mixture of thermal coherent states by coupling a thermal bath with a two-level system that is longitudinally coupled to a resonator. We find that the state of the resonator is a special mixture of two oppositely displaced thermal coherent states, whereas the two-level system remains thermal. This observation is verified by evaluating the second-order correlation coefficient for the resonator state. Moreover, we reveal the potential benefits of employing the mixture of thermal coherent states of the resonator in quantum thermometry. In this context, the resonator functions as a probe to measure the unknown temperature of a bath mediated by a two-level system, strategically bridging the connection between the two. Our results show that the use of an ancillary-assisted probe may enhance the precision and broaden the applicable temperature range.
	\end{abstract}	
 %--------------------------------------------------------------------------------
	\maketitle
 %--------------------------------------------------------------------------------
 \section{Introduction}
 %--------------------------------------------------------------------------------
 Quantum coherence is a fundamental concept in quantum mechanics related to the superposition of quantum states. Quantum coherence is a crucial resource that underpins many potential advantages of quantum technologies over classical~\cite{Madsen2012, PhysRevA.80.022304,PhysRevA.84.053802,PhysRevLett.59.2153,Takeda2013,Yokoyama2013,Ourjoumtsev2006}. Quantum coherence is utilized as a resource to improve the performance of several tasks, such as quantum information processing ~\cite{bouwmeester2000physics,RevModPhys.77.513}, quantum sensing and metrology~\cite{agarwal2012quantum,TAYLOR20161,ou2007multi,RevModPhys.90.035005,PhysRevLett.129.013602,ullah2023lowtemperature}, quantum computing~\cite{Andrew}, and energy harvesting ~\cite{PhysRevA.81.062346,mohseni2008environment}. In the context of thermodynamics, quantum coherence was shown to be a promising candidate for optimal charging of quantum batteries ~\cite{PhysRevLett.128.140501,PhysRevLett.127.100601,PhysRevE.102.052109,PhysRevLett.122.047702,PhysRevLett.120.117702,PhysRevLett.118.150601,PhysRevE.87.042123}, which enhances the efficiency of heat engines and thermal machines~\cite{pnas.1110234108, Hardal2015,ozgurEnt,BarisPRA,Hammam_2022}.
 
Coherent states are minimal uncertainty states that describe the maximum coherence and classical behavior~\cite{PhysRev.131.2766}. These states play a pivotal role in quantum physics, serving as fundamental tools for understanding quantum phenomena, quantum information processing, and the exploration of quantum optics~\cite{klauder1985coherent,zhang1990}.
%Superpositions of coherent states can be obtained in the resonant Jaynes-Cummings model if the cavity field is initially coherent with a large mean photon number~\cite{PhysRevLett.65.3385}.%
A thermal coherent state generalizes the notion of a coherent state to finite temperatures. By definition, it is a state that is produced by a thermalization operation applied to a displaced (coherent) state~\cite{Bishop1987, zhang1990, DTSVogt,PhysRevA.76.054307}. A few physical models have been proposed to prepare thermal coherent states (TCSs); for example, a superposition of displaced thermal states can be obtained via the resonant interaction of a two-level atom with a cavity field~\cite{PhysRevA.75.032114}. The TCS can be generated by quantum state engineering using an on-off detector scheme by pumping a nonlinear crystal with the help of a pump beam~\cite{Magaña-Loaiza2019}. An incident signal is a thermal field mixed with a vacuum field that results in conditional TCS. A probabilistic quantum amplifier can also produce a TCS using a thermal noise source and a photon number subtraction scheme~\cite{Usuga2010,PhysRevA.81.022302}. TCSs have been extensively studied in thermofield theory. The formalism of thermofield theory permits a more comprehensive formulation of the uncertainty relations, which include the effects of both quantum and thermal fluctuations~\cite{tfd,AmannCTS}. The TCSs can be particularly interesting as they saturate the generalized uncertainty relations, as regular coherent states do for the Heisenberg uncertainty relations. Recently, displaced thermal baths have been found to play a significant role in generating steady-state entanglement between two nitrogen-vacancy centers in a diamond host on an ultrathin yttrium iron garnet strip~\cite{kamran2022}.

Gaussian states are continuous variable states that exhibit the Gaussian probability in phase space and are fully characterized by their mean values and covariance matrices. %The most important Gaussian state is a vacuum state, which is also the eigenstate of the annihilation operator, whereas the thermal state is considered the most fundamental Gaussian state. 
Gaussian states in continuous-variable systems play a key role in the progress of quantum information and quantum metrology because of simple analytical tools and readily available experimental setups~\cite{RevModPhys.84.621, PhysRevA.98.012114}. Their versatility makes them suitable for tasks involving secret quantum key distribution and quantum simulation~\cite{TF,e17096072, PhysRevLett.94.020505}. Recently, continuous variable Gaussian systems have attracted much attention for performing low-temperature estimation tasks~\cite{PhysRevA.98.012114, PhysRevA.89.032128, PhysRevLett.128.040502, mirkhalaf2023operational}. 
 It is worth noting that, in temperature estimation, finding quantum Fisher information (QFI) and optimal measurements is very demanding\cite{PhysRevResearch.6.L032048}. It requires different techniques that depend on the scheme, system dynamics, and the parameter of interest. Yet, one can easily find them for the Gaussian systems. In multiparameter Gaussian quantum metrology, quantum Cramer-Rao bounds can still be saturated asymptotically for the estimation of multiparameters encoded in the multimode Gaussian states~\cite{PhysRevA.98.012114, PhysRevA.89.032128, Cenni2022thermometryof}. 

Quantum thermometry using a single qubit as a thermal probe represents a pivotal model with distinctive significance and practical applications~\cite{Boeyens_2023,Mehboudi_2019, Abiuso_2024}. However, single-qubit probes are inherently limited to estimating only a single temperature. To overcome this limitation and estimate multiple temperatures, a qubit probe can be coupled with a network of ancilla qubis~\cite{ullah2023lowtemperature,Zhang2022}. However, such networks demand precisely identical qubits with specific engineered anisotropic interactions~\cite{PhysRevE.99.042121}. Therefore, presenting readily available physical setups, like the coupled qubit-resonator model for quantum thermometry, becomes crucial, demonstrating potential feasibility for experimental implementation~\cite{RevModPhys.85.623}. This coupled system retains the same ability to estimate multiple temperatures precisely, hence enhancing the temperature estimation range.

In this paper, we address the following question: Can a harmonic oscillator (a multilevel system) in a non-thermal state—specifically, a displaced thermal state—outperform a two-level non-thermal probe in thermometry? To answer the aforementioned question, we propose a thermometry scheme based on a qubit-resonator system, where the resonator serves as the thermal probe. Our theoretical model prepares a mixture of two oppositely displaced TCSs, which are single-mode Gaussian states. The proposed model consists of a two-level system (qubit) coupled longitudinally to a resonator~\cite{Abari_2022,doi:10.1126/science.aao1511, PhysRevX.9.021056, Rouxinol_2016,Santos2019,
Wollack2022}. In addition, the qubit is also coupled to a thermal bath. The weak coupling with the bath results in the Gibbs thermal state for the qubit-resonator system. While the qubit remains in a thermal state, the steady state of the resonator yields a special mixture of thermal coherent states (MTCSs) with the same but opposite phase. To validate the coherence and thermal nature of the resonator state, we analyze the state's photon statistics with the help of a zero-time second-order correlation coefficient. In this scheme, the qubit encodes the temperature information of the thermal bath into the resonator state. A similar process of encoding a qubit into the energy levels of an oscillator is a subject of significant interest. Particularly, Gottesmann-Kitaev-Preskill states are utilized as an error correction technique~\cite{GKP2001}, and this approach extends across various quantum platforms to enhance resilience against thermal noise~\cite{PhysRevLett.128.170503, PhysRevA.93.012315, Shi_2019, SciPostPhysLectNotes.72, PhysRevA.108.052413}. A recent experiment investigated the generation of a quantum superposition of displaced thermal states within a microwave cavity using only unitary interactions with a transmon qubit~\cite{yang2024hot}.

We take advantage of the MTCSs of the resonator and investigate their role in temperature estimation. We use the resonator as a probe to estimate the temperature of a thermal bath. Due to the Gaussian nature of the probe state, we can analytically calculate the error propagation and QFI as relevant figures of merit. The thermometer (resonator) allows us to probe more than one temperature with higher precision compared to using a single thermalized harmonic oscillator. It is important to highlight that the precision enhancement obtained using the MTCSs of the resonator doesn't require degeneracy in the excited states of the probe~\cite{indP, Campbell_2018} or the need for an external control drive~\cite{Glatthard2022bendingrulesoflow, Mukherjee2019}. This makes the current scheme more practical in its precision improvement compared to those accounting for degenerate energy levels, which can be cumbersome to manage~\cite{indP, Campbell_2018}. We further investigate suboptimal measurements, in particular position and momentum measurements. The relative temperature error can be observed with high precision using quadrature measurements. The current setup closely resembles that in Ref.~\cite{PhysRevE.90.022102}, with the key distinction being that the qubit, in their case, acts as a working fluid interacting with both a hot and a cold bath. Additionally, the qubit is coupled to a resonator (referred to as a piston in their terminology). The primary objective of their work is to explore quantum states as thermodynamic resources, aiming to use them as effective tools for controlling and enhancing the efficiency of heat machines.

The rest of the paper is organized as follows: In Sec.~\ref{model}, we provide an overview of the model system and discuss its practical implementation. The results of this study are presented in Sec.~\ref{results}, where we discuss the preparation of the resonator state, which is the mixture of two oppositely displaced TCSs, and characterize its properties using the second-order correlation function. In Sec.~\ref{QT-TCS}, we discuss the advantageous application of the MTCSs that can enhance and broaden the range of thermometry precision. Finally, we conclude the paper in Sec.~\ref{conc}. The discussion on the relative error in temperature estimation is given in Appendix~\ref{ep} and the analytical expression of QFI is given in Appendix~\ref{appE}.
%---------------------------------------------------------------------------------
\section{The Model}\label{model}
%---------------------------------------------------------------------------------
  %-----------------------------------Fig1----------------------------------------
\begin{figure}[b!]
    \centering
    \includegraphics[scale=0.21]{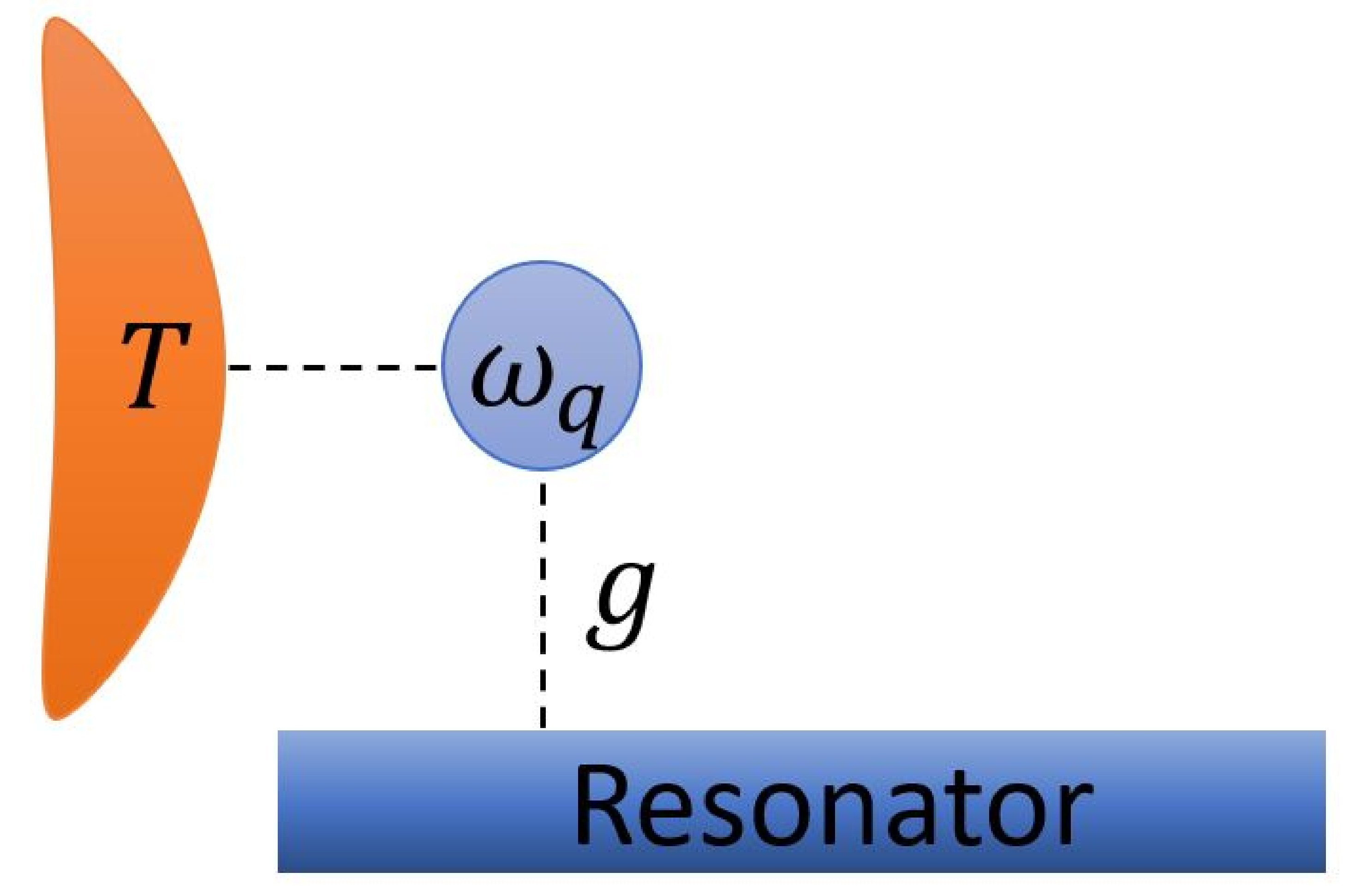}
    \caption{Model description. We consider a TLS; for example, in the current study, a qubit of frequency $\omega_q$ is coupled longitudinally to a resonator of frequency $\omega_R$. In addition, the qubit is also coupled to a thermal bath of temperature $T$. The qubit-resonator coupling is represented by $g$. We will observe the thermal effects of the bath in the resonator via qubit.}
    \label{fig:model}
\end{figure}
     %------------------------------------Fig1----------------------------------------
We consider a two-level system (TLS), such as a qubit interacting longitudinally with a single-mode bosonic resonator field. In addition, the qubit is coupled to a thermal bath of temperature $T$ as depicted in Fig.~\ref{fig:model}. The Hamiltonian of the system of interest is given by (we work in the units where $\hbar=1=k_B$ throughout this study)~\cite{LaHaye2009, PhysRevA.105.012201, PhysRevB.93.134501, PhysRevLett.115.203601}
\begin{equation}
    \hat{H}_S=\frac{\omega_q}{2}\hat{\sigma}_z+\omega_R\hat{a}^\dagger\hat{a}+g\hat{\sigma}_z(\hat{a}+\hat{a}^\dagger)\label{eqn1},
\end{equation}
where $g$ is the coupling strength between the qubit and resonator, $\omega_q$ and $\omega_R$ are the frequencies of the qubit and resonator, respectively, and $\hat{\sigma}_i(i=1,2,3)$ are the Pauli operators of the qubit. The bosonic creation (annihilation) operator of the resonator is represented by $\hat{a}^\dagger(\hat{a})$. The model described in Eq.~(\ref{eqn1}) differs from the Rabi model and more closely resembles the optomechanical model, where the Pauli matrix  $\hat{\sigma_z}$ takes on a role analogous to that of the cavity number operator $\hat{n}=\hat{a}^\dagger\hat{a}$~\cite{PhysRevE.99.042121}.  While the optomechanical model could be applicable, it would likely be constrained by more restrictive frequency and coupling coefficients, making it potentially less suitable for low-temperature regimes.
The longitudinal coupling between the resonator and the qubit has been implemented in circuit QED~\cite{PhysRevB.93.134501, PhysRevLett.115.203601}. This coupling has also been realized for a superconducting flux qubit coupled off resonantly to a transmission line resonator, offering advantages such as enhanced qubit lifetimes, high-fidelity measurements, entanglement over longer distances, and the generation of important microwave photon states for quantum communication~\cite{PhysRevA.69.062320,RevModPhys.85.623}.
Another mechanism utilizes strain coupling between nitrogen-vacancy centers and a diamond mechanical nanoresonator to enable long-range spin-spin interactions~\cite{PhysRevLett.110.156402}. It is also utilized for reversible information transfer between superconducting qubits and ultracold atoms using a microwave resonator~\cite{PhysRevA.79.040304}.
This model has also been used to predict and describe phonon blockade in a nanomechanical resonator coupled to a Cooper-pair box (i.e., a charge qubit) for observing strong antibunching and sub-Poissonian phonon-number statistics~\cite{PhysRevA.93.063861,LaHaye2009}.

In our setup, the qubit is coupled to a multimode bosonic bath, which is modeled as ensembles of harmonic oscillators and described by the Hamiltonian
\begin{equation}
    \hat{H}_B=\sum_k\omega_k\hat{b}^\dagger_k\hat{b}_k,
\end{equation}
and the system-bath interaction has the following form
\begin{equation}\label{eq:Hsb}
    \hat{H}_I=\sum_k \gamma_k\hat{\sigma}_x(\hat{b}_k+\hat{b}^\dagger_k),
\end{equation}
where $\hat{b}(\hat{b}^\dagger)$ is the annihilation (creation) bosonic operator of the k-th bath modes with $[\hat{b}_k,\hat{b}^\dagger_{k^\prime}]=\delta_{k,k^\prime}$, and $\gamma_k$ refers to the coupling strength between the k-th bath mode and qubit.

%---------------------------------------------------------------------------------
\section{Results}\label{results}
\subsection{Preparation of a mixture of thermal coherent states}
In our scheme, the ancilla qubit is coupled to both a resonator and a thermal bath of temperature $T$. We assume that the qubit is weakly coupled to the thermal bath, allowing the entire qubit-resonator state to reach a Gibbs thermal steady state. This assertion can be verified by considering a master equation derived under the Born-Markov and secular approximations~\cite{PhysRevA.98.052123, PhysRevE.90.022102, ullah2023lowtemperature}. It is important to note that an inconsistent derivation of the master equation may not result in the Gibbs steady-state of the system~\cite{09500349214552211}. For instance, the derivation of the master equation in the {\it{local basis}} may not result in the correct thermal state of the system~\cite{09500349214552211, PhysRevA.98.052123}. 

We consider weak-system bath coupling of the form given in Eq.~(\ref{eq:Hsb}); under this weak system-bath interaction, the reduced steady state of the qubit-resonator system is given by
\begin{equation}\label{ss}
    \hat{\rho} = \frac{1}{\mathcal{Z}(T)}e^{-\beta \hat{H}_S},
\end{equation}
where $\beta=1/T$ is the inverse temperature of the thermal bath and $\mathcal{Z}(T)=\text{Tr}[e^{-\beta \hat{H}_S}]$ is the partition function.
%To bring the global thermal state $\tilde{\rho}_{G}$ to the local basis, we first employ the back unitary transformation given in Eq.~(\ref{eq:unitary}). Subsequently, we will apply partial trace over the qubit state to get the desired state of the resonator in the local basis, which reads as
%\begin{equation}
%\hat{\rho}^{(R)}=Tr_q[\hat{\text{U}}\Tilde{\rho}_G\hat{\text{U}}^\dagger]=Tr_q[\hat{\text{U}}\Tilde{\rho}^{(q)}_{th}\otimes\Tilde{\rho}^{(R)}_{th}\hat{\text{U}}^\dagger].   
%\end{equation}
We are only interested in the reduced state of the resonator. Hence, performing the partial trace over the qubit (see Appendix~\ref{Der} for the detailed calculations), the resulting state of the resonator takes the following form
\begin{equation}\label{eqR}
\begin{aligned}
   \hat{\rho}^{(R)}&=pe^{-\theta(\hat{a}^\dagger-\hat{a})}\hat{\rho}^{(R)}_{th}e^{\theta(\hat{a}^\dagger-\hat{a})}\\
    &+(1-p)e^{\theta(\hat{a}^\dagger-\hat{a})} \hat{\rho}^{(R)}_{th}e^{-\theta(\hat{a}^\dagger-\hat{a})},
    \end{aligned}
\end{equation}
where $p=1/(e^{\beta\omega_q}+1)$ is the probability associated with the thermal state of the qubit and $\hat{\rho}^{(R)}_{th}$ is the thermal state of the resonator, such as $\hat{\rho}^{(R)}_{th}=\mathcal{Z}^{-1}_{th}(T)\sum_{n}e^{-\beta\omega_R\hat{a}^\dagger\hat{a}}|n\rangle\langle n|$ with $\mathcal{Z}_{th}(T)=\text{Tr}[e^{-\beta\omega_R\hat{a}^\dagger\hat{a}}]$ is the partition function. In Eq.~(\ref{eqR}), the state $ \hat{\rho}^{(R)}$ of the resonator is a statistical mixture of two oppositely displaced thermal coherent states with the same but opposite phase $\theta$. The density matrix obtained for the resonator (see Eq.~(\ref{eqR})) has two terms, each in the form of a thermal coherent state $\hat{D}(\alpha)\hat{\rho}_{th}\hat{D}^\dagger(\alpha)$ with displacement operator defined as $\hat{D}(\alpha)=e^{-(\alpha^*\hat{a}-\alpha\hat{a}^\dagger)}$ and $\alpha$ is in general a complex number. However, in our case, it is real and given by $\theta=g/\omega_R$.
Generally, a thermal coherent state is obtained if a thermal state is displaced by an amount of $\alpha$ ($\theta$ in our case) in phase space. On the other hand, if we trace out the resonator, we find that the two-level system is in a thermal state, represented by the density matrix given below
\begin{equation}
\hat{\rho}^{(q)} =\frac{1}{\mathcal{Z}_q}
\begin{pmatrix}
e^{-\beta\omega_q/2} &  \\
0 & e^{\beta\omega_q/2}
\end{pmatrix},
\end{equation}
where $\mathcal{Z}_q=e^{-\beta\omega_q/2}+e^{\beta\omega_q/2}$ is the partition function. This result is expected since the qubit is weakly coupled with the bath with energy exchange interaction given by $\hat{H}_{I}$, and longitudinally coupled to the resonator, satisfying the relation $[\hat{\sigma}_z,\hat{H}_S]=0$.
To validate our analytical results, we have made a numerical comparison and uploaded the Python code to GitHub~\cite{Asghar2024} (we refer the readers to Appendix~\ref{comp} for comparison of our analytical results with the numerical simulations).
%-----------------------------------Figure 2----------------------------------------------
\begin{figure*}[t!]
    \centering
    \subfloat[]{
    \includegraphics[scale=0.53]{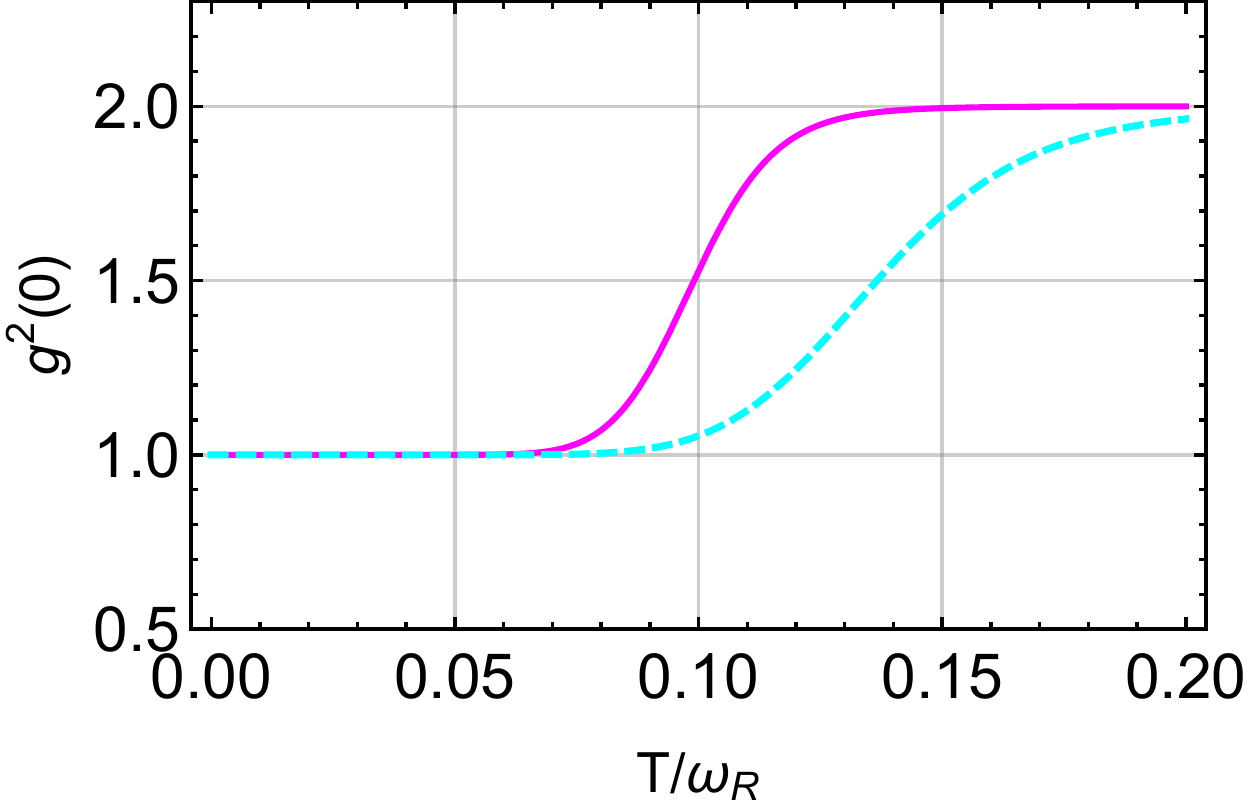}}
    \subfloat[]{
    \includegraphics[scale=0.465]{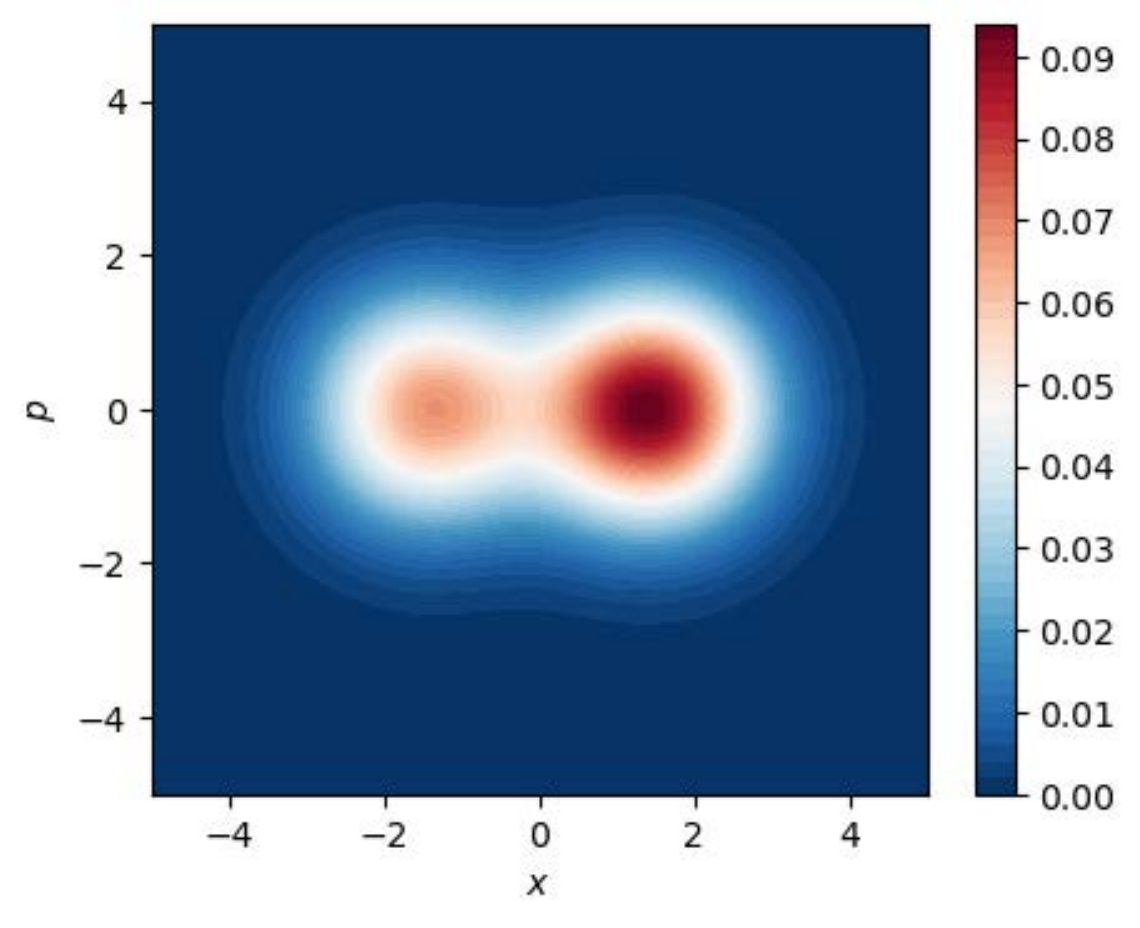}}
    \subfloat[]{
    \includegraphics[scale=0.465]{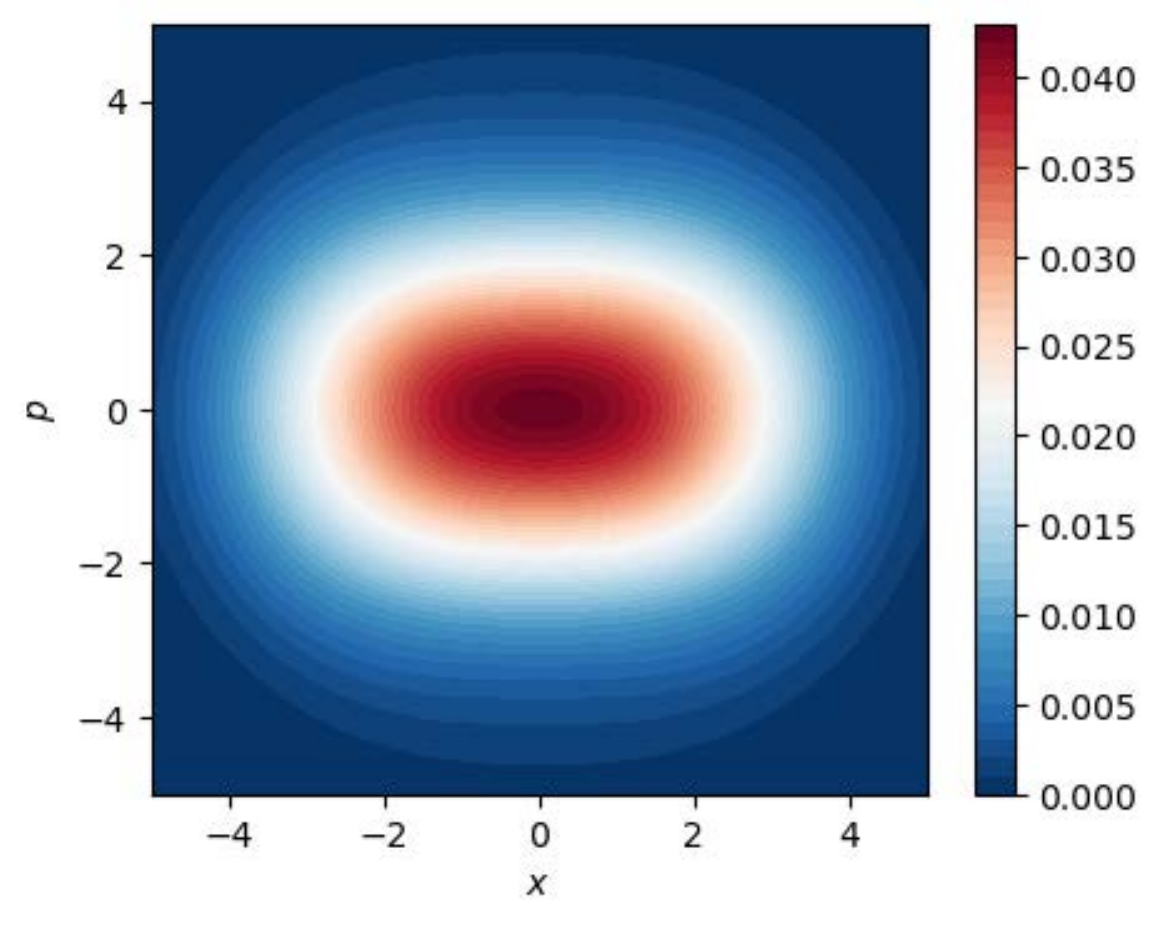}}
    \caption{ \textbf{(a)} The second-order correlation coefficient $g^2(0)$ for the steady state of the resonator as a function of bath temperature $T$. The solid magenta and cyan dashed curves are for $g=0.01$ and $g=0.04$, respectively. The other parameters are $\omega_q=1$ and $\omega_R=1$. In \textbf{(b)} and \textbf{(c)}, we plot phase diagrams using the Wigner function for the state of the resonator at two different temperature limits: $T=0.03$ (low-$T$) and $T=2$ (high-$T$). The other parameters are set to $\omega_q=0.01$, $\omega_R=1$, and $g=1$. All the system parameters are scaled with $\omega_R$.}
    \label{fig:2}
\end{figure*}
%-----------------------------------Fig2----------------------------------------------

\textit{Photon statistics:}
We employ a zero-time, second-order correlation function to 
investigate the coherence properties of the resonator's steady state. The second-order correlation function provides information on the sub-Poissonian and super-Poissonian statistics of the photon field, and it is given by~\cite{vogel2006quantum,PhysRevA.105.053701}
\begin{equation}
g^2(0)=\frac{\langle\hat{a}^\dagger\hat{a}^\dagger\hat{a}\hat{a}\rangle}{(\langle\hat{a}^\dagger\hat{a}\rangle)^2}.
\end{equation}
After calculating the average photon number $\langle \hat{n}\rangle=\langle\hat{a}^\dagger\hat{a}\rangle$ and $\langle\hat{a}^\dagger\hat{a}^\dagger\hat{a}\hat{a}\rangle$, the second-order correlation coefficient evaluates to~\cite{PhysRevA.105.053701}
\begin{equation}\label{corr}
    g^2(0)=\frac{\theta^4+2\Bar{n}^2+4\theta^2\Bar{n}}{(\Bar{n}+\theta^2)^2},
\end{equation}
where $\Bar{n}=({e^{\omega_R/T}-1})^{-1}$ represents the thermal average number of excitations (or thermal occupation number) for a given mode of the resonator. Photons in a coherent state with a Poisson distribution yield $g^2(0)=1$. Sub-Poission (antibunched photons) and Super-Poisson (bunched photons) statistics of photons give $g^2(0)<1$ and $g^2(0)>1$, respectively. Photonic states with sub-Poisson statistics are strictly nonclassical. Photons in a thermal state have $g^2(0) = 2$, and hence distributions with $g^2(0)>2$ and $1<g^2(0)<2$ are called super-thermal and sub-thermal (or partially coherent) distributions.

For numerical convenience, all system parameters in this study have been rescaled by the resonator frequency $\omega_R$, making them dimensionless unless otherwise mentioned. We examine both the low and high temperature limits relative to the resonator frequency $\omega_R$. We plot $g^2(0)$ as a function of bath temperature $T$ for different values of the coupling strength $g$ in Fig.~\ref{fig:2}. The results suggest that at lower temperatures $T\ll 1$, the state of the resonator adheres to Poissonian statistics. However, as the temperature of the thermal bath increases, the degree of coherence in the resonator decreases. Another factor affecting the distribution is the coupling strength between the resonator and qubit; stronger coupling results in a slower decrease in the coherence as bath temperature increases.

At low temperature, $T\approx 0$, $p\sim 1$, where $p$ represents the occupation probability of the ground state of the qubit. Consequently, we can ignore the qubit's excited state. In this scenario, the resonator state can be approximated to a single vacuum-displaced state, which is confirmed by the second-order coherence function value $g^2(0)=1$ at low temperature in Fig.~\ref{fig:2}.  However, with the increase in temperature of the thermal bath, coherence decreases because the phase of the oppositely displaced thermal state negatively affects the coherence in the resonator. At high temperatures (when we refer to the high $T$ limit, it implies $T\gg \omega_R$), the weights of oppositely displaced thermal states in Eq.~(\ref{eqR}) become approximately the same $p=1-p\sim 1/2$. Accordingly, the resonator state can be approximated to a thermal state because the coupling strength $g$ cannot be arbitrarily large (which can inject coherence into the resonator). We confirm this thermalization through the behavior of the second-order coherence function at high temperatures, which approaches the value $g^2(0)=2$, as depicted in Fig.~\ref{fig:2}. The second-order coherence function shows that the thermal state behavior is dominating the coherent character in the high temperature region. 

To further illustrate the effect of temperature on thermalization, we employ the Wigner function, a quasiprobability distribution that provides insight into the phase-space structure of the state. Unlike classical distributions, it can showcase non-classical features like quantum coherence and interference~\cite{walls2008quantum}. The Wigner function for an arbitrary density matrix $\hat{\rho}$ is given by (we consider $\hbar=1$)
\begin{equation}
    W(x, p) = \frac{1}{\pi} \int_{-\infty}^{\infty} \langle x - y | \hat{\rho} | x + y \rangle \, e^{-2ipy } \, dy,
\end{equation}
where $x$ and $p$ shows the position and momentum, respectively.
We plot the  Wigner function for the resonator state at relatively low and high temperatures in Figs.~\ref{fig:2}(b) and~\ref{fig:2}(c), respectively. At lower temperatures,  Fig.~\ref{fig:2}(b) depicts two oppositely displaced thermal coherent states with different weights. However, at higher temperatures (such as $T = 2$), these states are indistinguishable, indicating that the state approximates a thermal distribution (for more details on the Gaussian nature of the probe state, see Appendix~\ref{gauss} ).\\
%------------------------------------------------------------------------------
\textit{Multimode resonator:}
To investigate further, we consider a multimode resonator coupled to a qubit. The Hamiltonian of this is given by
\begin{equation}
    \hat{H}=\frac{\omega_q}{2}\hat{\sigma}_z+\sum_i\omega_i\hat{a}^\dagger_i\hat{a}_i+\sum_ig_i\hat{\sigma}_z(\hat{a}^\dagger_i+\hat{a}_i),
\end{equation}
where $g_i$ is the coupling strength between the qubit and each resonator mode of frequency $\omega_i$. In the case of identical resonator modes, where all the modes have the same frequency $\omega_i$ and coupling strengths $g_i$, the steady states for all modes have the same second-order correlation coefficient  $g^2_1(0)=g^2_i(0)$. We now assume that all the modes have different frequencies and coupling strengths.
%--------------------------------------Fig4-------------------------------------------
\begin{figure}[t!]
    \centering
    \includegraphics[scale=0.6]{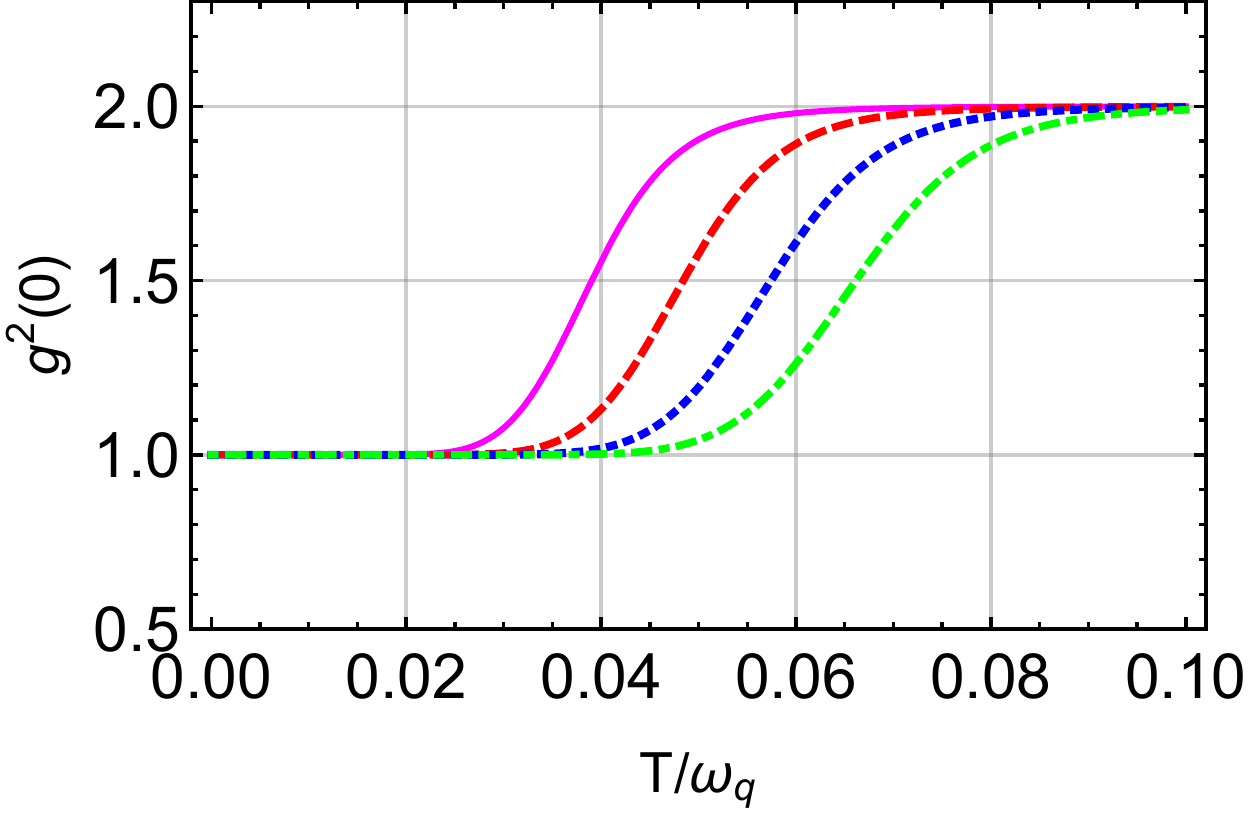}
    \caption{Second-order correlation coefficient $g^2(0)$ as function of bath temperature $T$ for different values of frequencies $\omega_i$: $\omega_1=0.3$ (solid magenta line), $\omega_2=0.4$ (dashed red line), $\omega_3=0.5$ (dotted blue line), and $\omega_4=0.6$ (dot-dashed green line). The other parameters are $\omega_q=1$ and $g_1=g_2=g_3=g_4=0.01$. All the system parameters are scaled with $\omega_R$. In this plot, all the system parameters are scaled with qubit frequency $\omega_q$.}
    \label{fig3}
\end{figure}  
%---------------------------------------Fig4------------------------------------------
The second-order correlation coefficient for the resonator having i-th modes can be written as
\begin{equation}
g_i^2(0)=\frac{\theta^4_i+2\Bar{n}^2_i+4\Bar{n}_i\theta^2_i}{(\Bar{n}_i+\theta^2_i)^2}.
\end{equation}
 We plot the second-order correlation coefficient $g^2(0)$ for various resonator modes as a function of bath temperature in Fig.~\ref{fig3}. We assume identical coupling strengths but different frequencies of the modes. In the case of a multimode resonator, the effects of all modes become evident when they occupy distinct energy states, indicating that each resonator has a unique frequency. 

\section{Thermometry with the MTCS\texorpdfstring{\lowercase{s}}{s}}\label{QT-TCS}
In this section, we investigate the MTCSs of the resonator as a probe for temperature estimation. To ensure clarity, we revisit fundamental concepts in quantum parameter estimation. We exploit the Gaussian formalism to calculate QFI using the covariance matrix of the resonator probe and compare it with the classical Fisher information obtained from more practical observables by measuring either the position or momentum of the resonator.
%We propose feasible observables by measuring position and momentum operators using the resonator's Gaussian state. Finally, we exploit the Gaussian formalism to calculate QFI and classical Fisher information of this state from the covariance matrix and compare the precision obtained by these two.
Finally, employing the error propagation method, we calculate the relative error in temperature, which is an alternative method for precision estimation (see Appendix~\ref{ep} for details).
\subsection{Fundamental concepts in thermometry}\label{T-con}
To begin with, we review some fundamental concepts that are used in quantum thermometry. While temperature is a parameter, not a quantum observable, it can be estimated through appropriate measurements on the probe, which in our scheme is the resonator. The choice of measurement influences the accuracy of estimation; for example, in the most general case, the positive operator value measure (POVM) and an estimator $\hat{T}$ constructed from the measurement data turn these measurements into temperature estimation. Therefore, we will use the local parameter estimation theory to quantify the precision of estimation~\cite{paris2009quantum}. A central quantity in parameter estimation theory is the QFI, which has a lower bound on the fluctuations of any unbiased estimator. The fluctuations in such estimators obey the Cram\'{e}r-Rao bound, such as~\cite{Helstrom1969, PhysRevLett.72.3439}
\begin{equation}\label{CRB}
    \Delta T\ge1/\sqrt{n\mathcal{F}_c(T)},
\end{equation}
where $n$ is the number of measurement outcomes and $\mathcal{F}_c(T)$ is the classical Fisher information with respect to the unknown parameter $T$ and associated with the particular choice of measurement and is given by
\begin{equation}
    \mathcal{F}_c(T)=\int dx p(x| T)[\partial_T\ln p(x|T)]^2,
\end{equation}
where $\partial_T:=\partial/\partial T$ is the partial derivative with respect to $T$ and $p(x| T)$ denotes the conditional probability for an output measurement x given the unknown parameter $T$. The Cram\'{e}r-Rao bound in Eq.~(\ref{CRB}) can be saturated in the asymptotic limit by selecting a maximum likelihood estimator~\cite{NEWEY19942111}. The fundamental task is to find an ultimate bound on precision. This can be obtained by finding the measurement that can minimize Eq.~(\ref{CRB}). Therefore, the corresponding precision is quantified by QFI, and it can be defined as~\cite{Cenni2022thermometryof}
\begin{equation}
\mathcal{F}_Q(T):=\max_{s}\mathcal{F}_C(T)=Tr[\rho_T L_T^2],
\end{equation}
where $s$ stands for a specific measurement performed and $L_T$  is the hermitian symmetric logarithmic derivative (SLD) operator defined by the equation $2\partial_T\rho_T=L_T\rho_T+\rho_T L_T$. The measurements attaining QFI can be optimal if the estimation error has a lower bound, which is given by the quantum Cram\'{e}r-Rao lower bound inequality 
\begin{equation}
\Delta T\ge 1/\sqrt{n\mathcal{F}_Q(T)}.
\end{equation}
Therefore, the maximum information about an unknown parameter obtained from the repeated measurements of state $\rho_T$ can be quantified by QFI. 
\subsection{Results}
%-------------------------------------------Figure5----------------------------------------%
\begin{figure}[t!]
    \centering
    \includegraphics[scale=0.6]{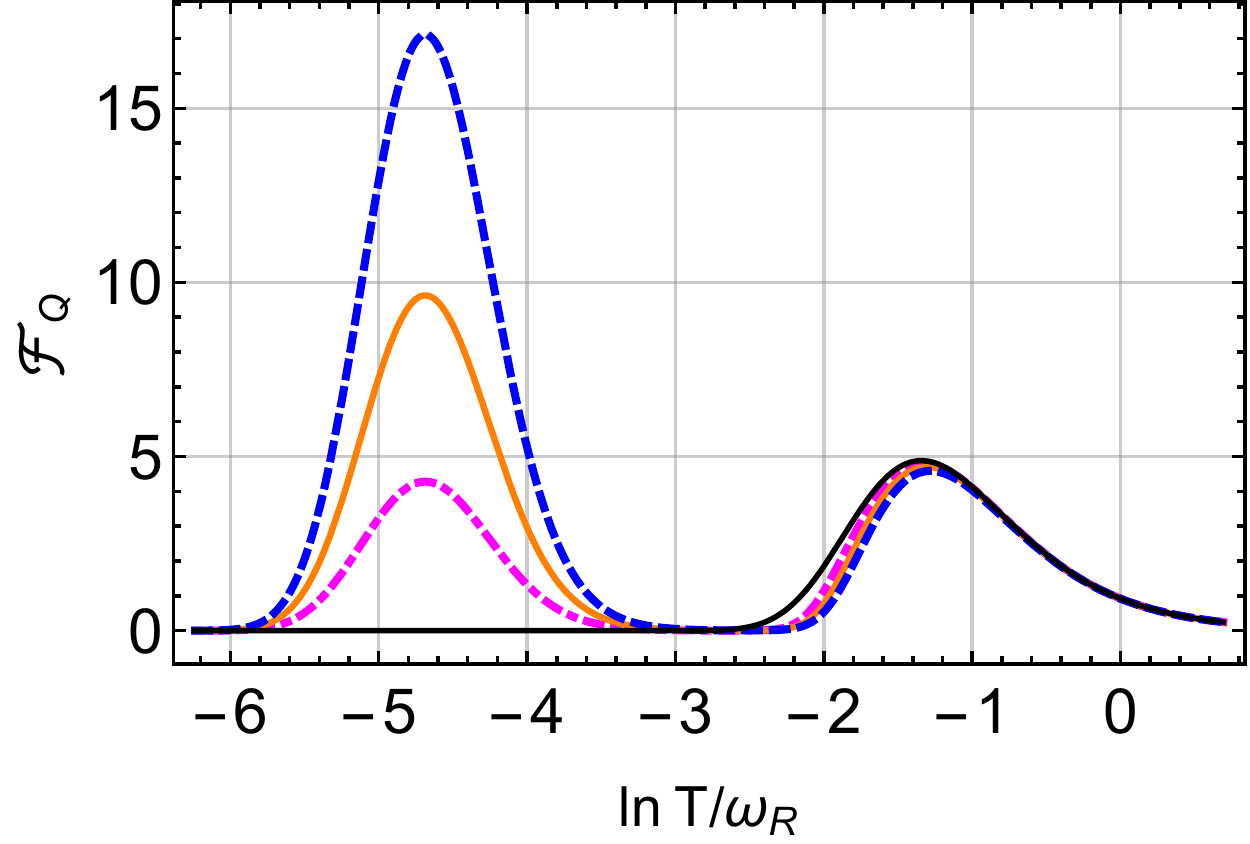}
\caption{ QFI $\mathcal{F}_Q$ for the MTCSs of the resonator plotted against the bath temperature $T$ for different values of coupling strength $g$. The dot-dashed (magenta), solid (orange), and dashed (blue) lines are for $g=0.02$, $g=0.03$, and $g=0.04$, respectively. We have added the QFI (solid black line) of a harmonic oscillator of frequency $\omega_\text{osc}$ at thermal equilibrium for comparison. The other parameters are $\omega_{R}=1$, $\omega_q=0.04$, and $\omega_\text{osc}=1$. All the system parameters are scaled with $\omega_R$.}.\label{fig4}
\end{figure}
%-------------------------------------------Figure5----------------------------------------%
\textit{Quantum Fisher information:}
To quantify the information about $T$ contained within the MTCSs, as given in Eq.~(\ref{eqR}), we will calculate the QFI.
Since Eq.~(\ref{eqR})  represents a single-mode Gaussian state, we can utilize the Gaussian formalism for the analytical calculation of QFI. We carefully select these parameters to ensure the system consistently remains in the Gaussian state for the parameter ranges considered in this study (see Appendix~\ref{gauss} for details). In addition, we avoid the strong coupling regime due to the absence of suitable experimental systems. In general, a Gaussian state is completely characterized by its first- and second-order canonical moments.
The first moment gives the expectation value of a canonical coordinate using state \( \hat{\rho}\), and is defined as $d_j = \langle \hat{\chi}_j \rangle_{\hat{\rho}}$.
Here $\hat{\chi}^{\text{T}}=(\hat{x}, \hat{p})$ is a vector which consists of position $\hat{x}=(\hat{a}+\hat{a}^{\dagger})/\sqrt{2}$ and momentum-like $\hat{p}=i(\hat{a}^{\dagger}-\hat{a})/\sqrt{2}$ quadrature operators of the resonator. The second-order moments form the covariance matrix (CM), defined by
\begin{equation}\label{cm}
\sigma_{ij} = \langle \{ \hat{\chi}_i, \hat{\chi}_j \} \rangle_{\hat{\rho}} - 2\langle \hat{\chi}_i \rangle_{\hat{\rho}}\langle \hat{\chi}_j \rangle_{\hat{\rho}}.
\end{equation}
Information about the temperature of the bath is encoded into the CM of the resonator. The QFI based on this CM can be calculated via the expression~\cite{monras2013phase, PhysRevA.88.040102}
\begin{equation}
    \mathcal{F}_Q(T)=\frac{Tr[\boldsymbol{\sigma}{^{-1}}(\partial_T\boldsymbol{\sigma})\boldsymbol{\sigma}{^{-1}}(\partial_T\boldsymbol{\sigma})]}{2(1+\mu^2)}+2\frac{(\partial_T\boldsymbol{\mu})^2}{1-\mu^4},\label{qfi-cov}
\end{equation}
where $\mu=1/\sqrt{\text{Det}\boldsymbol{\sigma}}$ is the purity of the state. 
%A Gaussian state is fully described by two statistical moments, such as the displacement vector and the CM. As an example, a vacuum state with $\Bar{n}=0$ is a Gaussian state whose displacement vector is zero and CM $\boldsymbol{\sigma} = \mathbb{I}/2$, where $\mathbb{I}$ is a $2\times2$ identity matrix. 
In the case where the resonator probe with frequency \(\omega_{\text{osc}}\) is in a thermal state defined by temperature \(T\), the QFI can be calculated using Eq. (15) with the CM of the thermal state \(\boldsymbol{\sigma} = (2\bar{n} + 1)\mathbf{I}_2\). Here \(\bar{n}\) represents the mean photon number, given by \(\bar{n} = \text{Tr}(\hat{\rho} \hat{a}^\dagger \hat{a})\), and $\mathbf{I}_2$ is a $2\times 2$ identity matrix. Accordingly, the QFI for the thermal state of a harmonic oscillator is given by $\mathcal{F}^{HO}_Q=\omega_\text{osc}^2\text{csch}^2(\omega_\text{osc}/2T)/4T^4$. This expression is plotted in Fig.~\ref{fig4}  (black solid curve) for comparison with our ancilla-assisted scheme. As expected, a single peak appears in the QFI associated with the thermal state of the resonator.

We calculate the expectation values of the quadratures using resonator state and these are given below
\begin{equation}
    \begin{aligned}
        \langle\hat{x}\rangle=2\theta(1-2p), \quad \langle\hat{p}\rangle=0\\
        \langle\hat{x}^2\rangle= 2\Bar{n}+1+4\theta^2, \quad \langle\hat{p}^2\rangle=2\Bar{n}+1.
    \end{aligned}
\end{equation}
Using these expectation values in Eq.~(\ref{cm}),  one can readily derive the covariance matrix for the MTCSs and it is expressed below as
\begin{equation}\label{tcs-cov}
    \boldsymbol{\sigma}=\left(
\begin{array}{cc}
 2 \Bar{n}+1+16 \theta ^2p(1-p) & 0 \\
 0 & 2\Bar{n}+1 \\
\end{array}
\right).
\end{equation}
Exploiting the formula of QFI (in Eq.~(\ref{qfi-cov})) and using the covariance matrix given in Eq.~(\ref{tcs-cov}), the final expression of QFI takes the following form:

\begin{equation}\label{qfi}
    \mathcal{F}_Q(T)=\frac{\text{A}_1+\text{B}_1}{2 T^4\omega_R^4},
\end{equation}
where $A_1$ and $B_1$  are introduced for brevity and explicitly given in Appendix~\ref{appE}.
%-------------------------------------------Figure7----------------------------------------%
\begin{figure}[b!]
    \centering
    \includegraphics[scale=0.29]{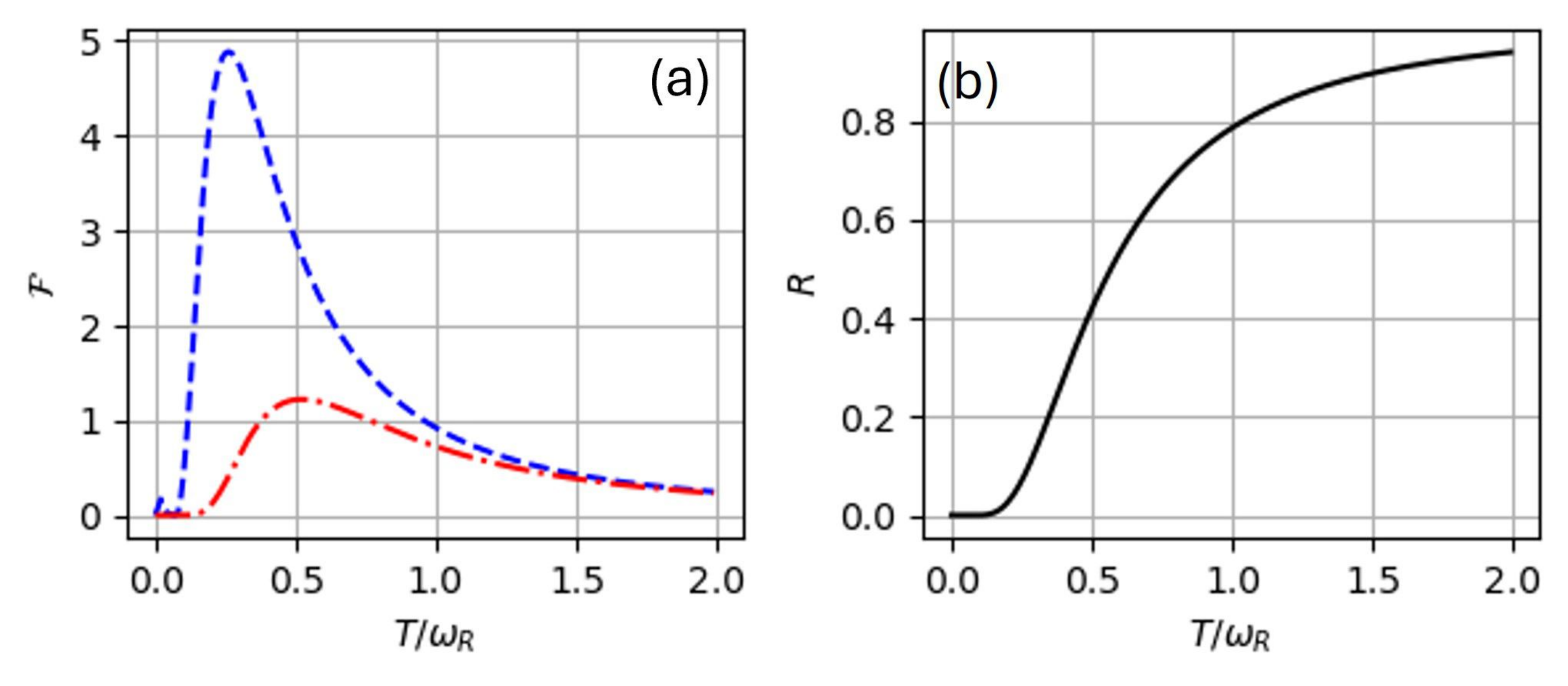}
\caption{ \textbf{(a)} The classical (red dot-dashed) and quantum (blue dashed) Fisher information as a function of $T$ with the coupling strength value $g=0.01$.\textbf{(b)} Fisher information ratio $\mathcal{R}=\mathcal{F}_C/\mathcal{F}_Q$ as a function of T. The plot is obtained by fixing the parameters to $\omega_q=\omega_R=1$ and $g=0.01$. All the system parameters are scaled with $\omega_R$.}
    \label{fig5}
\end{figure}
%-------------------------------------------Figure7----------------------------------------%
%We use the QFI given in Eq.~(\ref{qfi}) for the graphical explanation of our results obtained to characterize the precision of temperature estimation.
The QFI given in Eq.~(\ref{qfi}) is plotted as a function of the bath temperature $T$ for different values of coupling strength $g$ in Fig.~\ref{fig4}.
A notable advantage of our ancilla-assisted scheme is that by exploiting the MTCSs of the resonator, we can probe multiple temperatures with higher accuracy. In contrast, a single optimal sensing point exists in the QFI of a resonator probe (without ancilla qubit) thermalized with the bath, as shown in Fig.~\ref{fig4} (black solid curve). %To better comprehend the results, we plotted the QFI of a harmonic oscillator represented by a black solid curve for $\omega_{osc}=1$. This comparison highlights that a harmonic oscillator cannot probe more than one temperature. However, the resonator with the MTCSs can resolve more than one temperature. 
In our scheme, the upper peak in QFI (Fig.~\ref{fig4}) related to higher temperatures is almost indistinguishable from that of a harmonic oscillator at thermal equilibrium. However, the emergence of the lower peak is associated with the presence of the ancilla qubit. Consequently, the existence of the lower peak in the QFI requires the ancilla qubit to be on a lower energy scale, i.e., $\omega_q \ll \omega_R$.  %As a thermometer, the resonator is sensitive not only to its energy but also to the energy of the qubit to which it is coupled. This sensitivity increases with increasing coupling strength $g$. 
The height of the low-$T$ peak can be adjusted by varying $g$ while the peak's position can be shifted by changing the qubit frequency $\omega_{q}$. At near resonance, $\omega_q \sim \omega_R$, the lower peak is not present in QFI. In our scheme, the ancilla qubit acts as a transducer~\cite{PhysRevA.98.042124}, imprinting information about the temperature of the thermal bath onto the resonator state.

The ancilla qubit imprints information from the bath onto the resonator. The low-temperature peak in the QFI arises from the interaction between the ancillary qubit and the resonator, where the bath information is stored in the temperature-dependent coherence of the resonator. In contrast, the peak at higher temperatures is linked to the diagonal elements of the resonator. The amplitude of the lower QFI peak increases with the coupling strength $g$, and we observe a rapid increase in QFI as $g$ rises, as shown in Fig. 11 (see Appendix E). Additionally, it is possible to shift the location of the lower peak by tuning the qubit frequency $\omega_q$. In this way, the position and height of the lower peak can be precisely adjusted through appropriate tuning of $\omega_q$ and $g$.

 It is important to note that the current scheme does not require highly degenerate excited states to resolve multiple peaks in QFI, as in Ref.~\cite{Campbell_2018}, a task that can be challenging in practice. 
An alternative approach to obtaining multiple optimal sensing points in low-temperature estimation involves employing a dynamically controlled multilevel quantum probe in contact with the sample~\cite{Mukherjee2019}. 
%------------------------------Posittion QFI 8--------------------------------------------------%
 \begin{figure}[t!]
     \centering
     \subfloat[]{
     \includegraphics[scale=0.62]{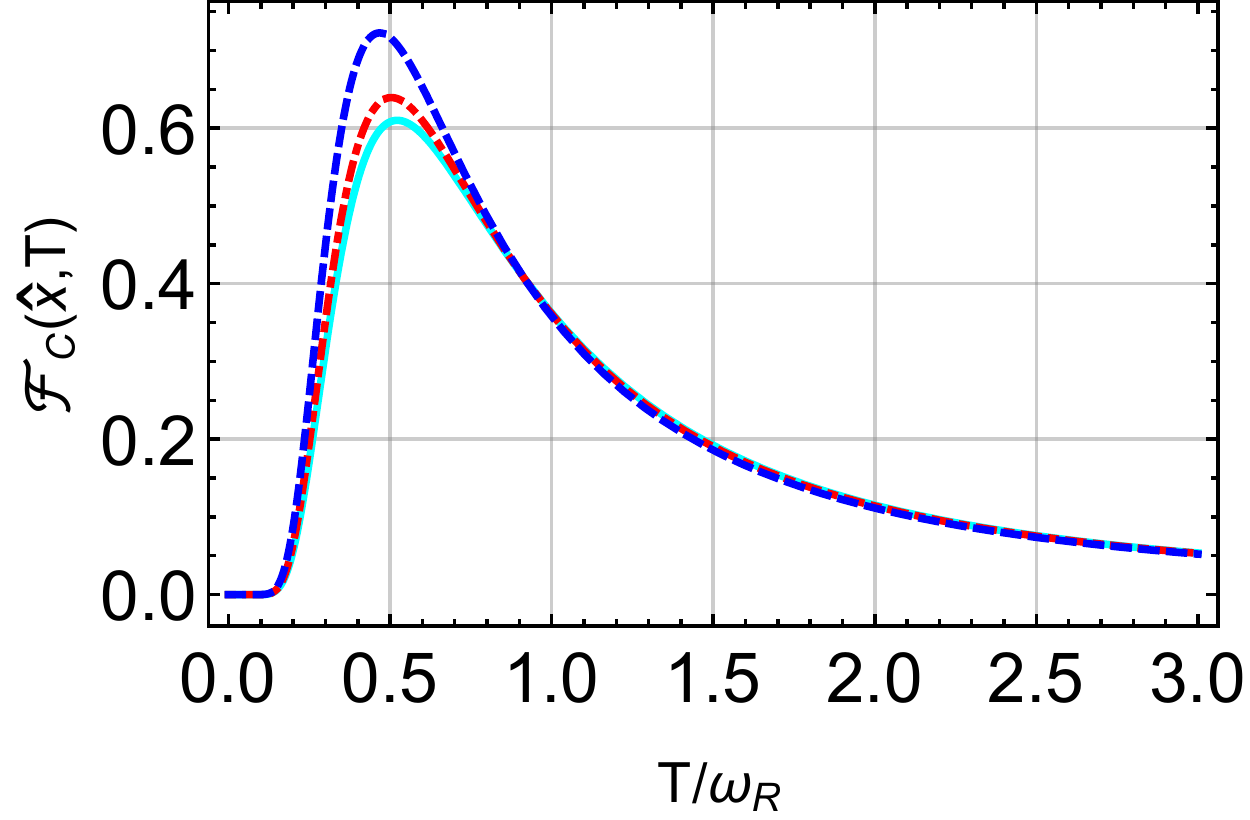}}\\
     \subfloat[]{
     \includegraphics[scale=0.6]{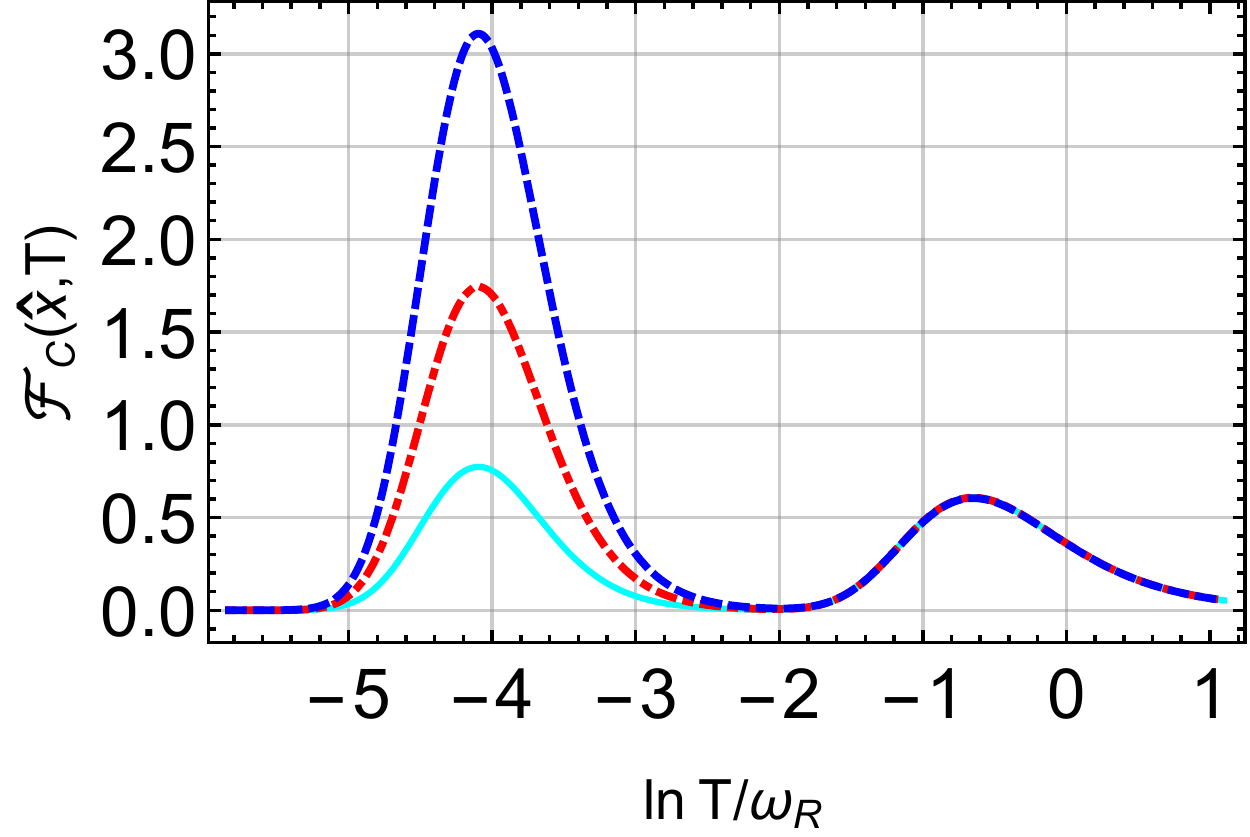}}
\caption{\textbf{(a)} CFI $\mathcal{F}_C(\hat{x}, T)$ as a function of $T$ for various values of coupling strength $g$ based on the quadrature measurement $\hat{x}$. Here, the solid (cyan), dot-dashed (red), and dashed (blue) lines are for $g=0.001$, $g=0.08$, and $g=0.15$, respectively. The other parameters are set to $\omega_q=1$ and $\omega_R=1$ . \textbf{(b)} $\mathcal{F}_C(\hat{x}, T)$ as a function of $T$ for various values of coupling strength $g$ such as the solid (cyan), dot-dashed (red), and dashed (blue) lines are for $g=0.02$, $g=0.03$, and $g=0.04$, respectively. The other parameters are $\omega_q=0.04$ and $\omega_R=1$. All the system parameters are scaled with $\omega_R$.}
\label{fig6}
 \end{figure}
%------------------------------Posittion QFI 8--------------------------------------------------%

\textit{Suboptimal thermometry and feasible measurements:}
One can also compute the classical Fisher information (CFI) associated with the Gaussian state measurement performed on the resonator probe.
We remark that the position and momentum quadratures belong to the Gaussian measurements and are characterized by the covariance matrix $\boldsymbol{\sigma}^k$ with $k\in\{\hat{x},\hat{p}\}$, which labels the measurement. 
Accordingly, the CFI for classical Gaussian distributions is well known and is given by~\cite{monras2013phase}
\begin{equation}\label{cf}
\begin{aligned}
\mathcal{F}_C(k,T)&= \partial_T\boldsymbol{d}^T(\boldsymbol{\sigma}+\boldsymbol{\sigma}^k)^{-1}\partial_T\boldsymbol{d}\\
&+\frac{1}{2}\text{Tr}\Big[\Big((\boldsymbol{\sigma}+\boldsymbol{\sigma}^k)^{-1}\partial_T\boldsymbol{\sigma}\Big)^2\Big],
\end{aligned}
\end{equation}
where $\boldsymbol{d}$ and $\boldsymbol{\sigma}$ represent the displacement vector and covariance matrix of the system, respectively. The covariance matrix for position measurement can be written as $\boldsymbol{\sigma}^x=\lim_{R\rightarrow\infty}(1/R,R)$ with some squeezing parameter $R\in[0,\infty]$~\cite{Cenni2022thermometryof}.
After some simple algebra, the simplified form of the CFI, $\mathcal{F}_{C}(\hat{x},T)$, based on position measurements, evaluates to
 \begin{equation}\label{eq:fc-cov}
\mathcal{F}_C(\hat{x},T)= \frac{|\partial_T\langle \hat{x}\rangle|^2}{\sigma_{11}}+\frac{|\partial_T\sigma_{11}|^2}{2\sigma_{11}^2}.
 \end{equation}
A similar expression can also be obtained for the measurement of the observable $\hat{p}$ using measurement covariance matrix $\boldsymbol{\sigma}^p=\lim_{R\rightarrow\infty}(R, 1/R)$. For our scheme, the analytical expression for CFI based on position measurement can be derived using Eqs.~(\ref{tcs-cov}) and (\ref{eq:fc-cov}), as given by
\begin{equation}\label{cfi}
\mathcal{F}_C(\hat{x},T)= \mathcal{F}_C(\hat{x},T)_1+\mathcal{F}_C(\hat{x},T)_2,
\end{equation}
where the two terms are given as
\begin{equation}
    \begin{aligned}
        \mathcal{F}_C(\hat{x},T)_1=\frac{\alpha}{T^4 \Big(4 \eta+\coth \left(\Phi\right)\Big)}, \text{and}\\
        \mathcal{F}_C(\hat{x},T)_2=\frac{\Big(2 \eta  \omega  \tanh \left(\phi\right)+\omega_R \text{csch}^2\left(\Phi\right)\Big)^2}{8 T^4 \Big(\eta +\coth \left(\Phi\right)\Big)^2}
    \end{aligned}
\end{equation}
respectively.
For the  sake of simplicity, we introduced the following parameters, given by
\begin{equation}
\eta =\theta ^2 \text{sech}^2\left(\phi\right),\quad \alpha=16 \theta ^2p^4 \omega ^2 e^{\frac{2 \omega }{T}},
\end{equation}
where $\phi=\omega_q/2T$ and $\Phi=\omega_R/2 T$ are defined for brevity.
We present a comparison between the QFI and suboptimal yet less challenging CFI as a function of temperature $T$ for a fixed value of coupling strength $g$ in Fig.~\ref{fig5}(a). Quite notably, $\mathcal{F}_Q>\mathcal{F}_C$ in the low-temperature regime. However, in the high-temperature region, the QFI converges to the CFI. This indicates that suboptimal measurements are not very effective in the low-temperature regime; however, for $T>1$, they are comparable.
%It can be concluded from the two expressions such as QFI (Eq.~(\ref{qfi})) and CFI (Eq.~\ref{cfi}) that the QFI surpasses the CFI, as denoted by $\mathcal{F}_Q>\mathcal{F}_C$ in the low-temperature regime.
In Fig~\ref{fig5}(b), we compute the Fisher information ratio $\mathcal{R}=\mathcal{F}_C/\mathcal{F}_Q$ as a function of bath temperature $T$. %We fix the parameters to $\omega_q=\omega_R=1$ and $g=0.01$.
As depicted in the figure, the ratio begins at zero, undergoes an increase, and eventually saturates at higher temperatures.
The trend in the ratio of Fisher information aligns with the characteristics of the state of the resonator. At a high temperature $T$, the state of the resonator exhibits more classical characteristics, transitioning towards a thermal state as indicated by the second-order coherence function shown in Fig.~\ref{fig:2}, leading to a dominance of classical features and, consequently, the QFI approaching the CFI.

Fig.~\ref{fig6} illustrates the CFI for position measurement  $\hat{x}$ as described in Eq.~(\ref{cfi}).  We present the CFI for two scenarios: when the qubit and resonator are in resonance and when they are off-resonant in Figs.~\ref{fig6}(a) and~\ref{fig6}(b), respectively.  
The CFI in Fig.~\ref{fig6}(a) based on the position measurements does not display two peaks for the temperature $T$.
Meanwhile, when we change the frequency of the qubit coupled to the bath, such as $\omega_q=0.04$ and $\omega_R=1$, the CFI has two peaks with respect to $T$ as shown in Fig.~\ref{fig6}(b). However, the peak at low-$T$ has less height as compared to the peak of QFI in Fig.~\ref{fig4}, for the same values of coupling strength $g$.
\begin{figure}[t!]
    \centering
    \includegraphics[scale=0.6]{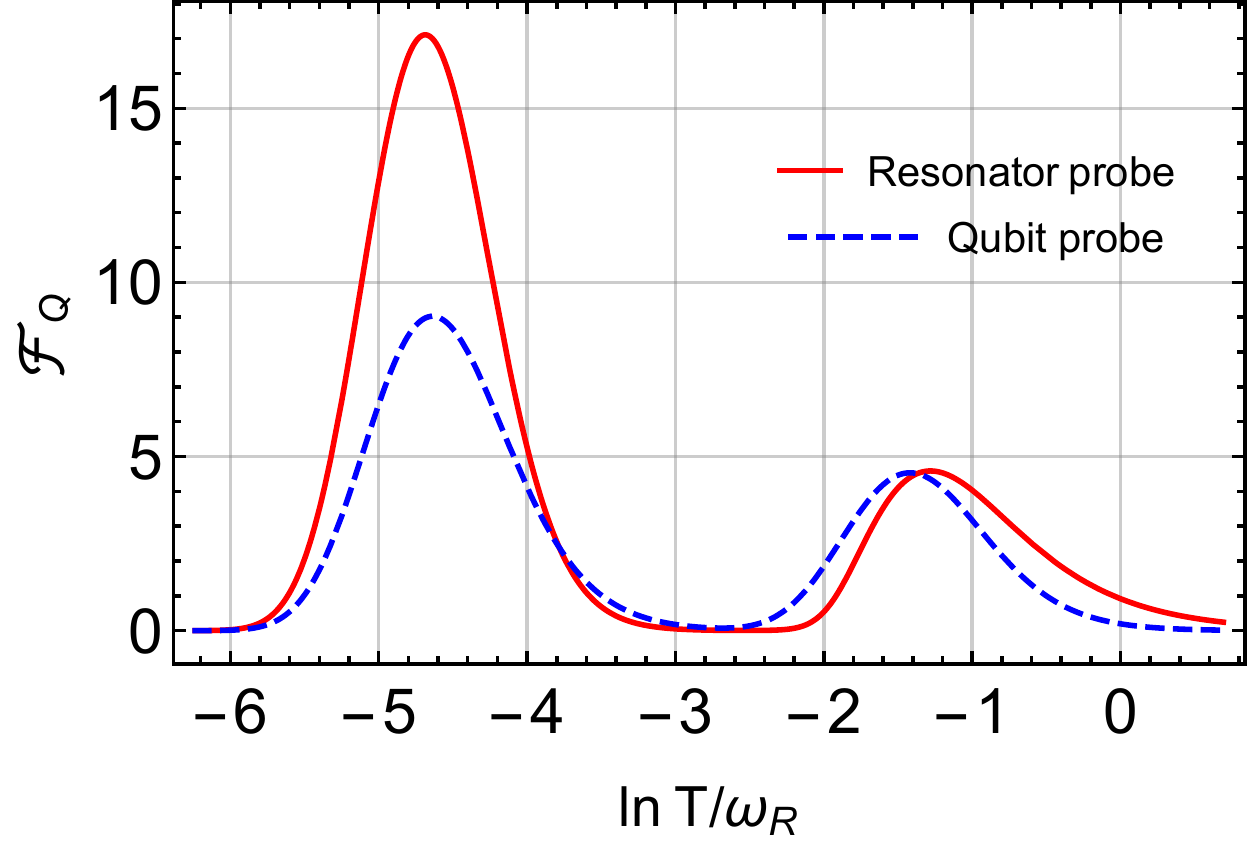}
    \caption{QFI is plotted as a function of temperature $T$ for two different probes. The blue dashed curve corresponds to the scheme presented in Ref.~\cite{ullah2023lowtemperature}, where a qubit is used as the thermometer. The parameters for this scheme are set to $\omega_a=0.04$ (ancilla qubit frequency), $\omega_p=1$, and $g=0.04$. The red curve represents the QFI for a probe resonator in the current work, with parameters $\omega_q=0.04$, $\omega_p=1$, and $g=0.04$. In both cases, all the parameters are scaled with the probe frequency $\omega_p$.}
    \label{QR}
\end{figure}
\subsection{Qubit vs Resonator probe}
The QFI for a two-level system or a harmonic oscillator probe in thermal equilibrium is generally similar. Simply adding more equally spaced energy levels—going from a two-level system to a multi-level system—doesn't substantially improve sensitivity. To enhance the performance of a thermometer, we investigated a non-thermal probe in this work, such as a resonator probe, which has the potential to outperform a qubit probe in temperature estimation. If we use qubit as a probe instead of a resonator, the QFI for such a qubit probe is given by~\cite{ullah2023lowtemperature}
\begin{eqnarray}\label{qfi-qubit}
	\mathcal{F}_{Q} = \frac{2\omega^2_p\text{sech}^2\big(\frac{\omega_p}{2T}\big)+\theta^2\omega^2_a\text{sech}^2\big(\frac{\omega_a}{2T}\big)}{8T^4},
\end{eqnarray}
where $\theta=\arctan(2g/\omega_p)$ and $\omega_p$ and $\omega_a$ are the transition frequencies of the probe and the ancilla qubit, respectively. We compare the results of the resonator probe with those of the qubit probe used in Ref.~\cite{ullah2023lowtemperature}, as shown in Fig.~\ref{QR}. In this Fig~\ref{QR}, the QFI is plotted as a function of bath temperature $T$: the dashed blue curve represents the QFI for the qubit thermometer (following Ref.~\cite{ullah2023lowtemperature}), while the solid red curve corresponds to the resonator probe. The results clearly indicate that, for the same set of parameters, the resonator probe's precision is nearly twice that of the qubit probe at low temperatures.  This finding contrasts with the QFI of a qubit and resonator in thermal equilibrium, where the difference is not significant.
This substantial improvement in precision underscores the practical advantages of the resonator probe, making it a more effective choice than a qubit-based thermometer.
\section{Conclusion and discussion}\label{conc}
%---------------------------------------------------------------------------------
We investigated the preparation of a resonator state, which is a special mixture of two oppositely displaced thermal coherent states, coupled to a thermal bath via an ancillary two-level system. In particular, our focus was on a longitudinally coupled qubit to a resonator, which serves as a representative model for a broad category of physical systems~\cite{{LaHaye2009, PhysRevA.105.012201, PhysRevB.93.134501, PhysRevLett.115.203601}}. The qubit is excited by attaching a thermal bath of temperature $T$ to it. The reduced state of the resonator is a special mixture of two oppositely displaced thermal coherent states with the same but opposite phase. To quantify its properties, we studied the photon statistics, and the results showed that the resonator state in Eq.~(\ref{eqR}) follows a Poissonian distribution ($g^2(0)\approx1$) at low temperatures, indicating the state's coherent properties. However, the photon statistics become super-Poissonian ($g^2(0)>1$) as the temperature of the bath increases, which confirms the thermal properties of a coherent state.

We exploit both the coherent and thermal nature of the MTCSs of the resonator and use it to estimate the unknown temperature of the thermal bath. For temperature estimation, we consider the resonator as a probe to measure $T$ with the assistance of a qubit attached to the thermal bath. Taking advantage of the Gaussian characteristics of the MTCSs, we employed tools from Gaussian quantum metrology to calculate QFI. Our results indicate that the state of the resonator, within an ancilla-assisted quantum thermometry scheme, not only enhances precision at low temperatures but also broadens the measurable range compared to using a single thermalized harmonic oscillator. Additionally, we compared the QFI with suboptimal yet more practical measurements based on either the position or momentum of the resonator. We report that while suboptimal measurements are unsuitable at low temperatures, QFI and CFI based on suboptimal position measurements are comparable at high temperatures.

 However, finding the optimal or best measurement to extract maximum information about temperature T using a resonator probe is a challenging task. The necessary and sufficient condition to estimate a parameter optimally is to construct the POVM by using the set of projectors over the eigenstates of the SLD. However, calculation of SLD is a non-trivial task, especially for coupled hybrid or continuous-variable systems. Moreover, the determination of POVM itself is not sufficient per se to fully answer this question. It is necessary to address the question of optimum estimator in addition to POVM, which can be searched as a function of the eigenvalues of the SLD. The latter task can be performed by classical post-processing of measurement results to saturate the Cram\'{e}r-Rao bound. \\
Experimental access to temperature-sensitive field quadratures and their second moments is feasible through techniques such as homodyne, heterodyne, and tomographic state reconstruction methods~\cite{Cenni2022thermometryof, PhysRevResearch.4.023191}. We can infer the temperature through straightforward quantum optical quadrature measurements~\cite{Raymer2004} on the cavity photons. The calculations show that the corresponding CFI ($\sim3$) is about 20$\%$ of the QFI ( $\sim 15$), as determined by comparing Figs.~\ref{fig4} and~\ref{fig6}. Therefore, while our exploration of suboptimal measurements holds practical relevance, our investigation into QFI highlights the fundamental potential of the system. Determining the exact optimal POVM for our system in order to fully utilize its precision potential for maximizing temperature information is an intriguing question with significant practical implications. If an experimental approach can be devised, possibly involving a combination of field quadrature measurements, it would be immensely valuable. However, due to the analytical complexities involved, we leave this avenue open for future research.

 Finally, our results may find applications in quantum thermometry, where the aim is to expand the applicable range of temperature estimation using quantum systems exposed to thermal baths. In addition, the low energy scale of the qubit with respect to the resonator can also be useful in broadening the range of temperature estimation in ancilla-assisted quantum thermometry schemes~\cite{PhysRevA.98.042124}.

\subsection*{Acknowledgment}
This work is supported by the Scientific and Technological Research Council of Turkey (TÜBİTAK) under grant number 122F371.
%------------------------------------------------------------------------------------
\appendix
%----------------------------------------------------------

%-------------------------------------------------------------------

%--------------------------------------------------------

\section{Derivation of the probe state}\label{Der}
In our thermometry scheme, which consists of a qubit-resonator system, the qubit is coupled to a thermal bath at temperature $T$. We assume that the system bath coupling is weak, allowing the qubit and resonator to reach a global Gibbs thermal state in the steady state. This condition holds under weak system-bath coupling and can be validated by using a global master equation within the Born-Markov and secular approximations, as detailed in references~\cite{Breuer,PhysRevA.98.052123, PhysRevE.90.022102, ullah2023lowtemperature}.
The reduced Gibbs thermal state of the qubit-resonator system can be written as~\cite{Breuer,lidar2020lecturenotestheoryopen}
\begin{equation}\label{gts}
    \Tilde{\rho}_G=\Tilde{\rho}_{th}^{(q)}\otimes\Tilde{\rho}_{th}^{(R)},
\end{equation}
where $\Tilde{\rho}_{th}^{(q)}$ and $\Tilde{\rho}_{th}^{(R)}$ are the qubit and resonator thermal density matrices, respectively. The thermal state of the qubit is given by
\begin{equation}
    \Tilde{\rho}_{th}^{(q)}=p|0\rangle\langle 0|+(1-p)|1\rangle\langle 1|,
\end{equation}
where 
\begin{equation}
    p=\frac{1}{e^{\beta\omega_q}+1}
\end{equation}
is the probability of qubit thermal state and the state of the resonator is
\begin{equation}
    \Tilde{\rho}_{th}^{(R)}=\sum_n\frac{e^{-\beta\omega_R\Tilde{a}^\dagger\Tilde{a}}}{\mathcal{Z}}|n\rangle\langle n|.
\end{equation}
The global thermal state of the qubit-resonator system is described in the global basis, but to focus on the resonator, we need to obtain its reduced local state. This requires the use of a transformation operator that allows us to switch from the global to the local basis, after which we can effectively trace out the qubit's degrees of freedom.
The transformation operator in this case is given by
\begin{equation}
    \hat{\text{U}}=e^{-\theta\hat{\sigma}_z(\hat{a}^\dagger-\hat{a})},
\end{equation}
where $\theta=g/\omega_R$. The system Hamiltonian in Eq.~(\ref{gts}) is expressed in the global basis, which is obtained through the application of the transformation operator. The explicit form of the Hamiltonian in the diagonal basis, after applying the transformation, is given by
\begin{equation}
    \Tilde{H}=\frac{1}{2}\omega_q\hat{\sigma}_z+\omega_R\Tilde{a}^\dagger\Tilde{a}-\frac{g^2}{\omega_R}.
\end{equation}
The frequencies of the qubit and bosonic field modes are unaffected by the longitudinal coupling. The transformed operators are given as
\begin{equation}
\begin{aligned}
    \Tilde{\sigma}^-=&\hat{\text{U}}\hat{\sigma}^-\hat{\text{U}}^\dagger=\hat{\sigma}^-e^{-\theta\hat{\sigma}_z(\hat{a}^\dagger-\hat{a})},\\
    \Tilde{a}&=\hat{\text{U}}\hat{a}\hat{\text{U}}^\dagger=\hat{a}-\theta\Tilde{\sigma}^+\Tilde{\sigma}^-.
    \end{aligned}
\end{equation}
The Gibbs thermal state can explicitly be written as
\begin{equation}
\begin{aligned}  
\Tilde{\rho}_G&=\left(p|0\rangle\langle 0|+(1-p)|1\rangle\langle 1|\right)\otimes\Tilde{\rho}_{th}^{(R)}.
\end{aligned}
\end{equation}
\begin{figure*}
    \centering
    \subfloat[]{
    \includegraphics[scale=0.21]{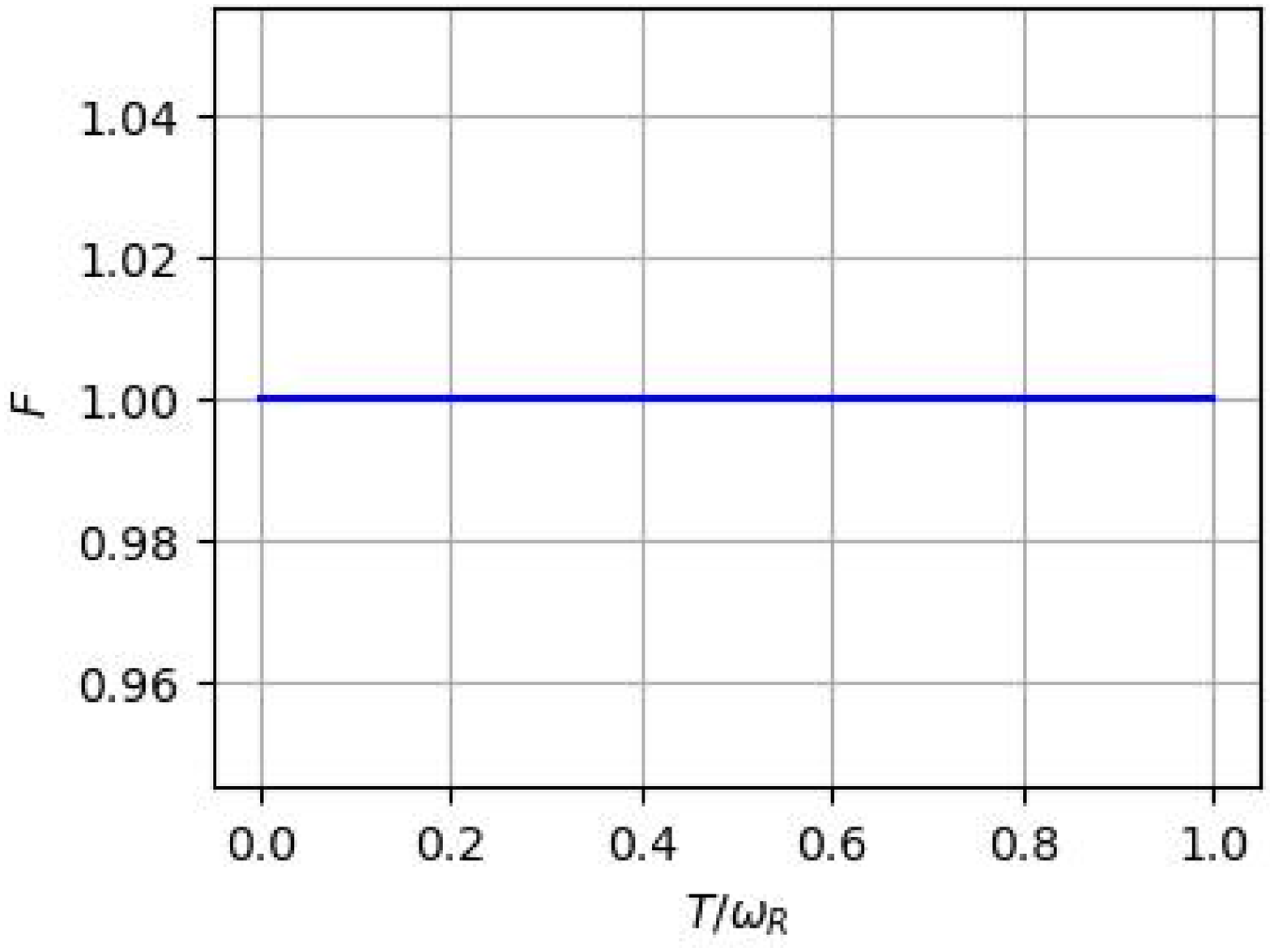}}
    \subfloat[]{
       \includegraphics[scale=0.57]{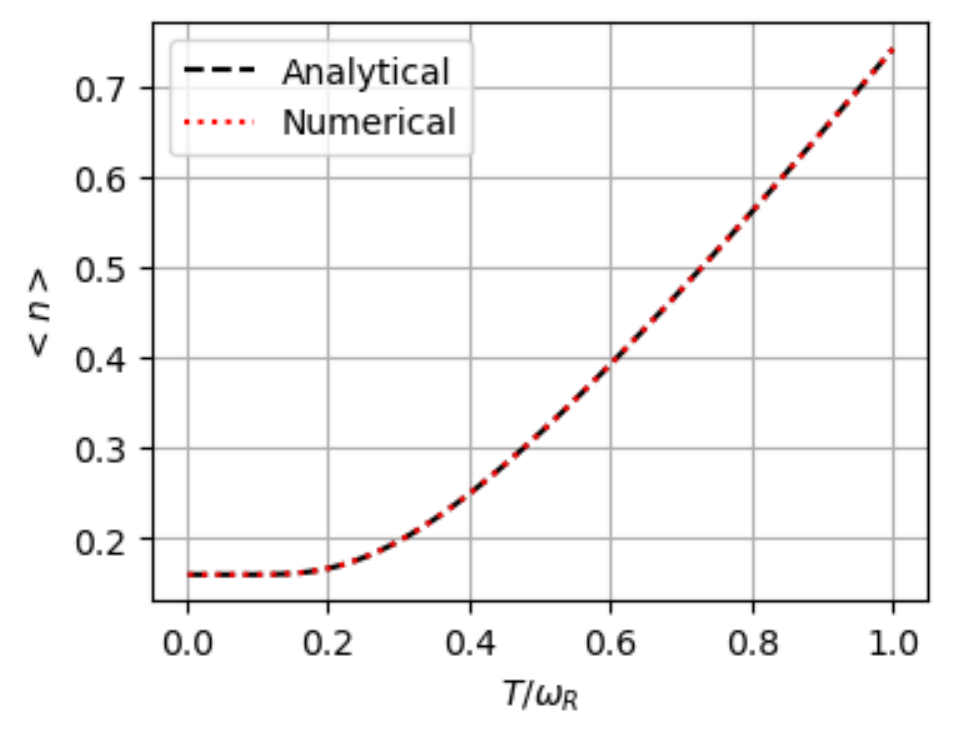}}
       \subfloat[]{
          \includegraphics[scale=0.57]{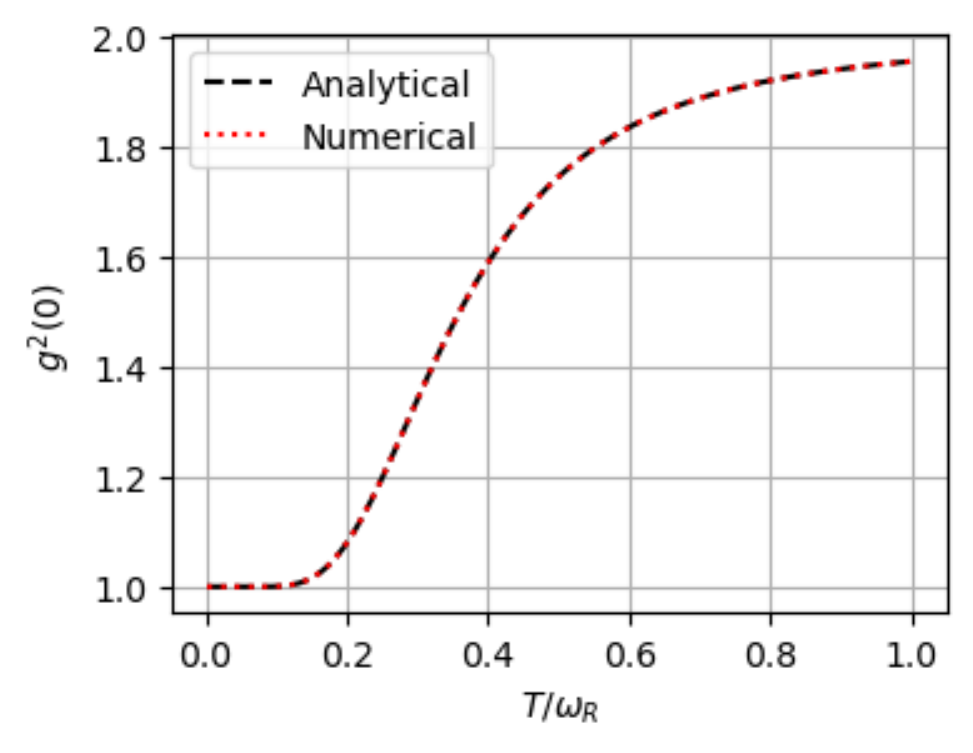}}
    \caption{\textbf{(a)} The fidelity is plotted as a function of temperature $T$ between the resonator state in Eq.~(\ref{eqR}) and the numerically obtained state. \textbf{(b)} The mean photon number $⟨n⟩$ for TCS with respect to
temperature T. The black dashed curve is plotted using Eq.~(\ref{mp}), and the red dotted curve is obtained with numerical simulations. \textbf{(c)} Represents the second-order coherence function $g^2(0)$ as a function of temperature $T$. The other parameters are $\omega_q=0.4$, $\omega_R=1$, and $g=0.4$. All the system parameters are scaled with $\omega_R$.}
    \label{fig7}
\end{figure*}
The unitary operator $\hat{\text{U}}$ can be written in expanded form as
\begin{equation}
    \hat{\text{U}} = 1 -\theta \hat{\sigma}_z (\hat{a}^\dagger - \hat{a}) + h.o.t.,
\end{equation}
where the notation h.o.t. stands for higher order terms in the expansion.
By first applying the unitary transformation and subsequently simplifying the resulting expression through the application of $\hat{\sigma}_z$ on the qubit states $|0\rangle$ and $|1\rangle$, we obtain
\begin{widetext}
\begin{equation}
    \begin{aligned}
     \hat{\rho}_G&=\hat{\text{U}}\left[\left(p|0\rangle\langle 0|\otimes\Tilde{\rho}_{th}^{(R)}+(1-p)|1\rangle\langle 1|\right)\otimes\Tilde{\rho}_{th}^{(R)}\right
     ]\hat{\text{U}}^\dagger\\
    &= \big[1-\theta\hat{\sigma_z}(\hat{a}^\dagger-\hat{a})+h.o.t.\big]\big[\big(p|0\rangle\langle 0|\otimes\Tilde{\rho}_{th}^{(R)}+(1-p)|1\rangle\langle 1|\big)\otimes\Tilde{\rho}_{th}^{(R)}\big]\big[1+\theta\hat{\sigma_z}(\hat{a}^\dagger-\hat{a})+h.o.t.\big]\\
    &=\big[1-\theta(\hat{a}^\dagger-\hat{a})+h.o.t.\big]\big[p|0\rangle\langle 0|\otimes\Tilde{\rho}_{th}^{(R)}\big]\big[1+\theta(\hat{a}^\dagger-\hat{a})+h.o.t.\big]\\
       &+\big[1+\theta(\hat{a}^\dagger-\hat{a})+h.o.t.\big]\big[(1-p)|1\rangle\langle 1|\otimes\Tilde{\rho}_{th}^{(R)}\big]\big[1-\theta(\hat{a}^\dagger-\hat{a})+h.o.t.\big]
    \end{aligned}
\end{equation}
We now apply partial trace over the qubit degrees of freedom that yields the state of the resonator
\begin{equation}
    \begin{aligned}
       \hat{\rho}^{(R)}
       &=\text{Tr}_q\big[1-\theta(\hat{a}^\dagger-\hat{a})+h.o.t.\big]\big[p|0\rangle\langle 0|\otimes\Tilde{\rho}_{th}^{(R)}\big]\big[1+\theta(\hat{a}^\dagger-\hat{a})+h.o.t.\big]\\
       &+\text{Tr}_q\big[1+\theta(\hat{a}^\dagger-\hat{a})+h.o.t.\big]\big[(1-p)|1\rangle\langle 1|\otimes\Tilde{\rho}_{th}^{(R)}\big]\big[1-\theta(\hat{a}^\dagger-\hat{a})+h.o.t.\big]\\
       &=\big[1-\theta(\hat{a}^\dagger-\hat{a})+h.o.t.\big]p\Tilde{\rho}_{th}^{(R)}\big]\big[1+\theta(\hat{a}^\dagger-\hat{a})+h.o.t.\big]\\
       &+\big[1+\theta(\hat{a}^\dagger-\hat{a})+h.o.t.\big]\big[(1-p)\Tilde{\rho}_{th}^{(R)}\big]\big[1-\theta(\hat{a}^\dagger-\hat{a})+h.o.t.\big]\\
       &=e^{-\theta(\hat{a}^\dagger-\hat{a})}p\Tilde{\rho}_{th}^{(R)}e^{+\theta(\hat{a}^\dagger-\hat{a})}+e^{+\theta(\hat{a}^\dagger-\hat{a})}(1-p)\Tilde{\rho}_{th}^{(R)}e^{-\theta(\hat{a}^\dagger-\hat{a})}
    \end{aligned}
\end{equation}
We can finally write the state of the resonator as
\begin{equation}
    \hat{\rho}^{(R)}=pe^{-\theta(\hat{a}^\dagger-\hat{a})}\hat{\rho}_{th}^{(R)}e^{+\theta(\hat{a}^\dagger-\hat{a})}+(1-p)e^{+\theta(\hat{a}^\dagger-\hat{a})}\hat{\rho}_{th}^{(R)}e^{-\theta(\hat{a}^\dagger-\hat{a})}
\end{equation}
\end{widetext}
\section{Analytical vs numerical results}\label{comp}
In this section, we present the results obtained by numerically solving Eq.~(\ref{ss}) and compare them with the analytical results derived from the resonator state in Eq.~(\ref{eqR}). The state of the resonator from Eq.~(\ref{eqR}) shows consistency with the numerically obtained state. It is verified by calculating the fidelity
\begin{equation}\label{fid}
    F(\hat{\rho}^{(R)},\hat{\rho}^\prime)=(\text{Tr}\sqrt{\sqrt{\hat{\rho}^{(R)}}\hat{\rho}^\prime\sqrt{\hat{\rho}^{(R)}}})^2
\end{equation}
between the two states, as shown in Fig.~\ref{fig7}(a). In above Eq.~(\ref{fid}), $\hat{\rho}^{(R)}$ is the analytical state of the resonator (Eq.~(\ref{eqR})) and $\hat{\rho}^\prime$ denotes the state obtained by numerically solving Eq.~(\ref{ss}).
To confirm our analytical results, we plot the mean photon number $\langle \hat{n}\rangle$  and second-order coherence function $g^2(0)$ as a function of T for the resonator state, as shown in Fig.~\ref{fig7}(b) and (c). The black dashed curves are plotted using the analytical expressions in Eq.~(\ref{mp}) and Eq.~(\ref{corr}), while the dotted points correspond to results obtained through numerical simulations using Eq.~(\ref{ss}). The numerical simulations are performed using open-source packages in QuTiP~\cite{JOHANSSON20121760} while the open source code is available on GitHub~\cite{Asghar2024}. We note a quantitative agreement (as evidenced by the overlap of both plots in the figures) between the two methods.
%-------------------------------------Fig5-------------------------------------%
\begin{figure}[t!]
    \centering
    \subfloat[]{
    \includegraphics[scale=0.14]{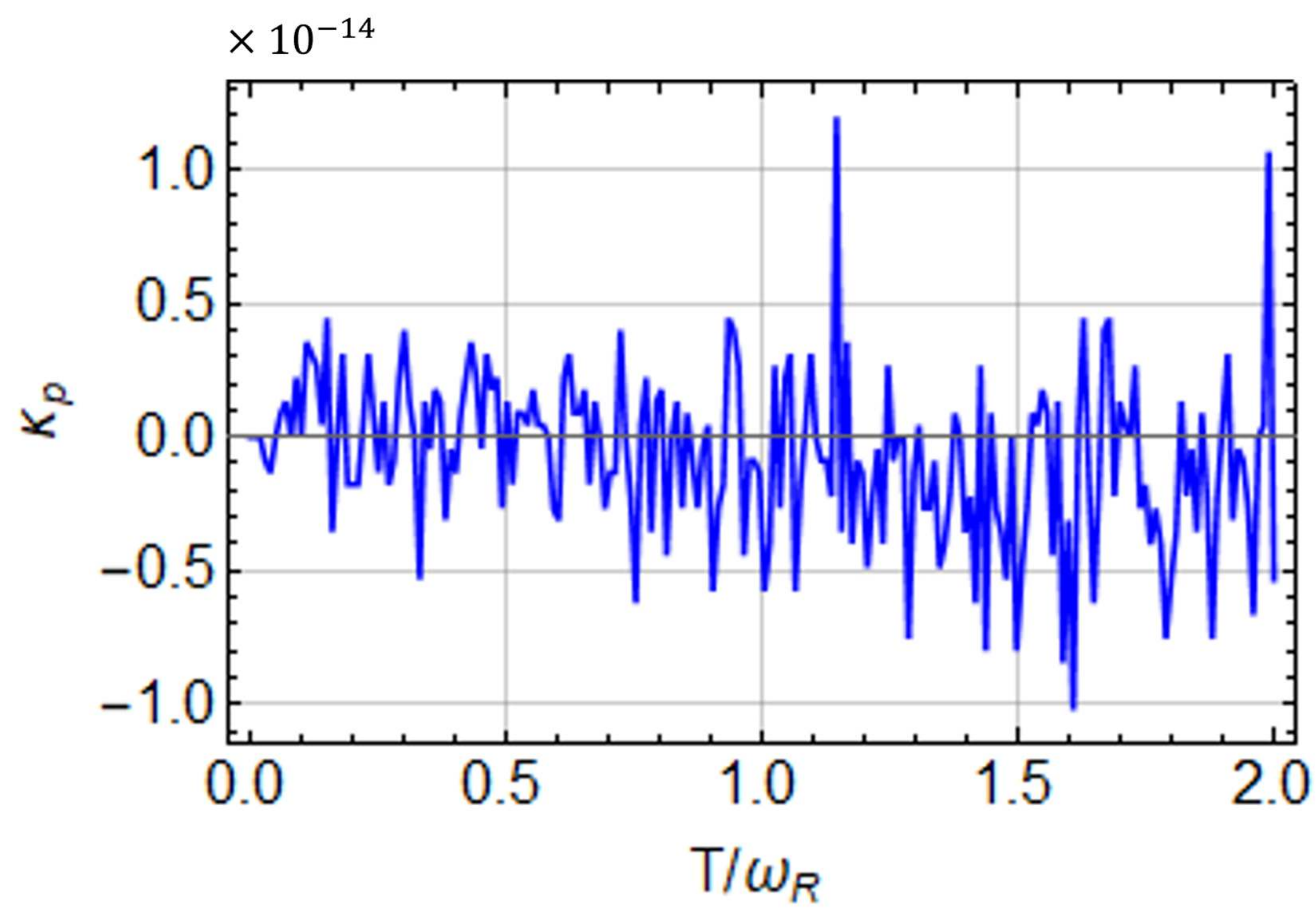}}\\
    \subfloat[]{
    \includegraphics[scale=0.14]{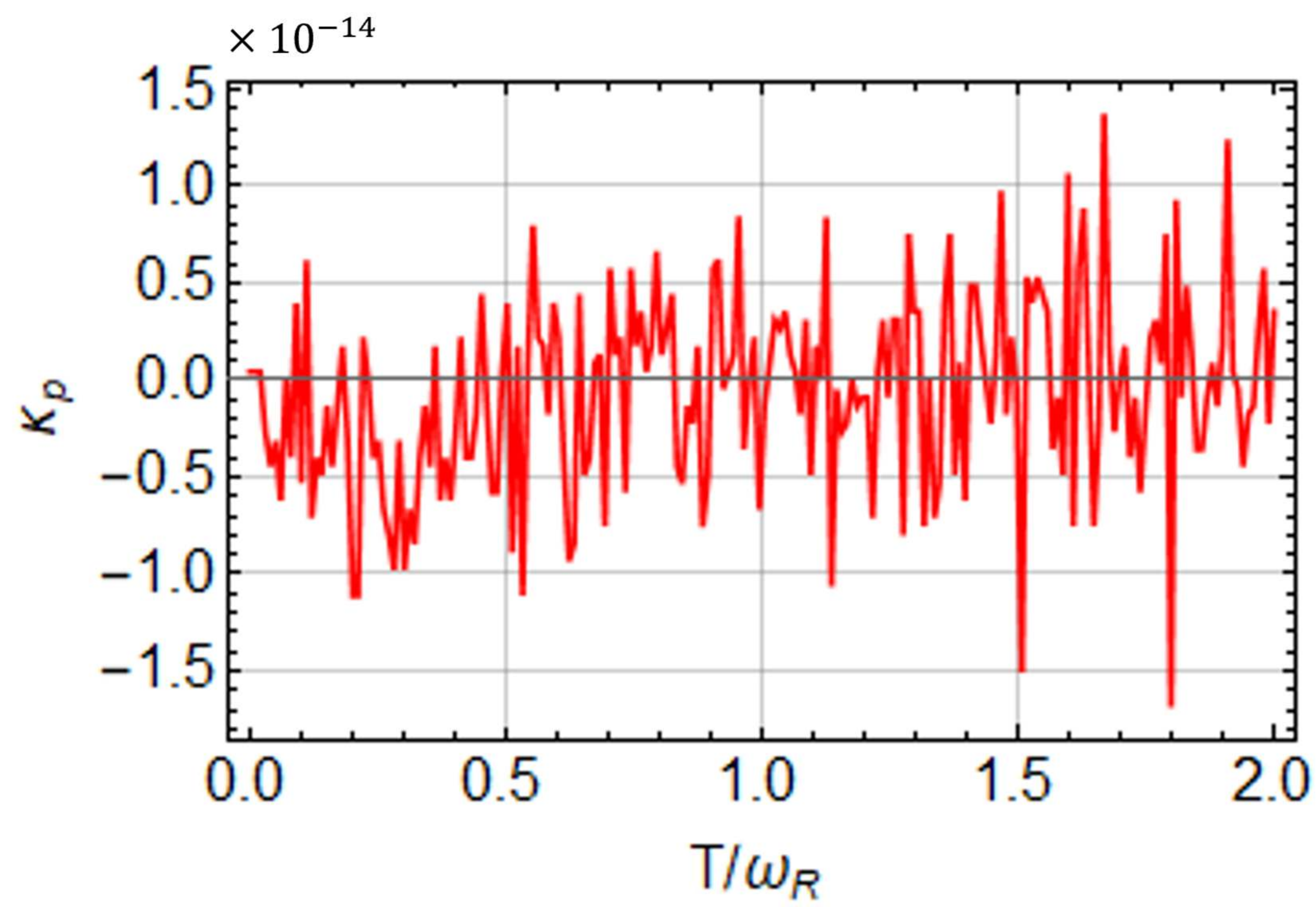}}
\caption{Excess kurtosis $\kappa_p$ (given in Eq.~(\ref{ek})) as a function of bath temperature $T$ for the state of the resonator given in Eq.~(\ref{eqR}). In \textbf{(a)}, the parameters are set to $\omega_q=0.04$, $\omega_R=1$, and $g=0.06$. In \textbf{(b)}, all parameters are identical to those used in Fig.~\ref{fig:2}(c). All the system parameters are scaled with resonator frequency $\omega_R$.}
    \label{fig8}
\end{figure}
%----------------------------------Fig5----------------------------------------%

%-------------------------------------------------------------%
\section{Gaussianity verification for the probe state\label{gauss}}
%-------------------------------------------------------------%
To quantify the Gaussian nature of the state of the resonator in Eq.~(\ref{eqR}), we calculate the excess kurtosis for a single quadrature, such as the phase quadrature $\hat{p}$ in our case. It is defined as~\cite{PhysRevLett.121.220501} 
\begin{equation}\label{ek}
    \kappa_p=\frac{\langle\hat{p}^4\rangle}{\langle\hat{p}^2\rangle^2}-3
\end{equation}
The excess kurtosis is a key indicator of non-Gaussianity, setting it apart from Gaussian statistics, where all higher-order cumulants beyond the second order are zero. An excess kurtosis of zero ($\kappa_p=0$) indicates Gaussian statistics, whereas non-zero values reveal deviations from Gaussian behavior. Specifically, a negative excess kurtosis ($\kappa_p<0$) signifies sub-Gaussian measurement statistics, while a positive excess kurtosis ($\kappa_p>0$) implies super-Gaussian statistics. The excess kurtosis $\kappa_p$ as a function of $T$ for the phase quadrature $\hat{p}$ is plotted in Fig.~\ref{fig8}(a). It is evident from the plot that, for a fixed coupling strength $g=0.06$, $\kappa_p$ remains close to zero throughout the temperature range considered. This observation indicates that the resonator state exhibits predominantly Gaussian characteristics under the specified parameters investigated in this study. Moreover, as discussed earlier in the main text, in the high-temperature limit ($T\gg\omega_R$) the weights of the oppositely displaced thermal coherent states become approximately equal, i.e., $p=1-p\sim 1/2$ in Eq.~(\ref{eqR}). To further clarify this assumption, we plot the excess kurtosis as a function of $T$ for $\omega_q=0.01$ and the strong coupling value of $g=1$ in Fig.~\ref{fig8}(b). The excess kurtosis for these parameters, which are identical to those in Fig.~\ref{fig:2}(c), remains unchanged, confirming the previous assumption that the state remains Gaussian. Thus, we conclude that the resonator state retains its Gaussian nature, validating the assumptions made in this study.

\section{Error propagation method}\label{ep}
We calculate the inverse of error propagation, which is an alternative way of characterizing precision. The error propagation to characterize the estimation error for any observable $\hat{X}$ is defined as~\cite{Cenni2022thermometryof}
\begin{equation}
    \delta^2 T_{EP}:=\frac{Var(\hat{X})}{|\partial_T\langle\hat{X}\rangle|^2}.
\end{equation}
%-------------------------------------Fig5-------------------------------------%
\begin{figure}[b!]
    \centering
    \includegraphics[scale=0.6]{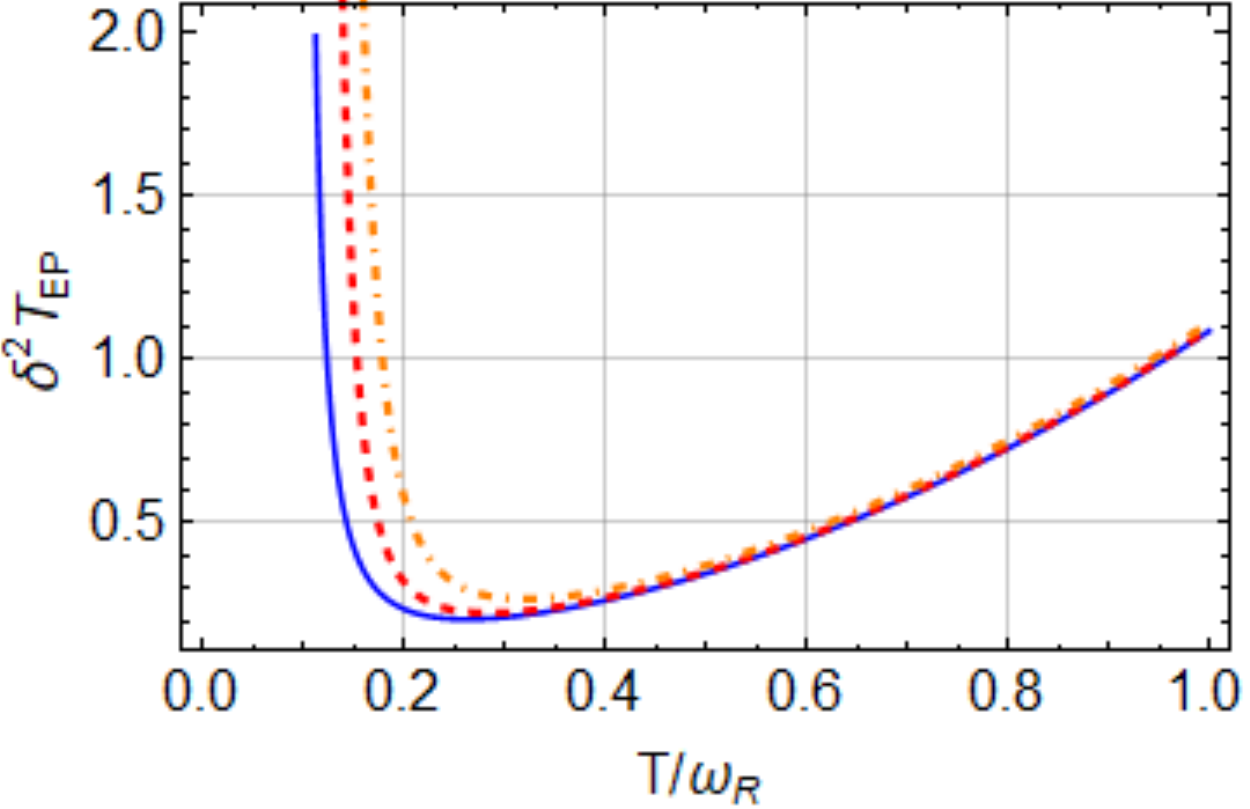}
\caption{ The uncertainty $\delta^2T_{EP}$ of $T$ as a function of bath temperature $T$ for different values of coupling strength $g$. The solid blue, red dashed, and orange dot-dashed curves correspond to $g=0.01$, $g=0.05$, and $g=0.1$, respectively. The other parameters are $\omega_q=1$ and $\omega_R=1$. All the system parameters are scaled with $\omega_R$.}
    \label{fig9}
\end{figure}
%----------------------------------Fig5----------------------------------------%
To estimate the temperature of the bath using the above formula, we need to find the expectation values $ \langle \hat{a}^\dagger\hat{a}\rangle$ and $\langle (\hat{a}^\dagger\hat{a})^2\rangle$. Employing Eq.~(\ref{eqR}), one can analytically derive these values. Then, the expectation values from Eq.~(\ref{eqR}) can easily be obtained, and these read as
\begin{equation}
\begin{aligned}
    \langle \hat{a}^\dagger\hat{a}\rangle=&\Bar{n}+\theta^2,\\
    \langle (\hat{a}^\dagger\hat{a})^2\rangle=&2\Bar{n}^2+\Bar{n}(4\theta^2+1)+\theta^4+\theta^2.\label{mp}
\end{aligned}
\end{equation}
The variance in the number operator of the MTCSs has the value
\begin{equation}
\Delta n^2=\Bar{n}(\Bar{n}+1)+(2\Bar{n}+1)\theta^2.
\end{equation}
From the above equations (i.e., Eq.~(\ref{mp})), it is evident that direct photon detection (with the measurement operator $\langle \hat{a}^\dagger\hat{a}\rangle$) can be employed for estimating the temperature $T$ of the thermal bath. By applying the error propagation formula, the uncertainty of $\delta^2 T_{EP}$ can be derived as follows:
\begin{equation}
  \delta^2T_{EP} =\frac{4 T^4 \sinh ^2\left(\Phi\right) \left[1+2 \theta^2 \sinh \left(2\Phi\right)\right]}{\omega_R^2}.
\end{equation}
The behavior of uncertainty $\delta^2 T_{EP}$ as a function of $T$ is shown in Fig.~(\ref{fig9}). This figure indicates that the uncertainty in T decreases as the coupling strength $g$ between the qubit resonator is decreased from $g=0.1$ (the orange dot-dashed curve) to $g=0.01$ (the blue curve). From this method, we can conclude that the temperature estimation is better within the weak coupling regime, which allows for more accurate temperature measurement.
\begin{widetext}
\section{Analytical expression of QFI}\label{appE}
The QFI of the resonator is analytically determined using the covariance matrix, yielding the expression:
\begin{equation}
    \mathcal{F}_Q=\frac{\text{A}_1+\text{B}_1}{2 T^4\omega_R^4},\quad \text{where} \quad \text{A}_1=\frac{4 \text{a}_1}{\text{a}_2+1}.
\end{equation}
In the above expressions, certain parameters are defined, given as follows:
    \begin{equation}
    \begin{aligned}
        \text{a}_1=\left[\frac{8 g^2 p^3 \omega_q  e^{\omega_q /T}/\Bar{n}+\Bar{n}^2 \omega_R^3 e^{\omega_R/T}}{2 \Bar{n}+1+16 \theta ^2p^2 e^{\omega_q /T}}\right]^2+\Bar{n}^2 \omega_R^6 e^{\frac{2 \omega_R}{T}} \quad,
        \text{a}_2=\frac{\tanh \left(\Phi\right)}{4 \eta+\coth \left(\Phi\right)}
         \end{aligned}
    \end{equation}
and
\begin{equation}
\begin{aligned}
        \text{B}_1=\frac{\tanh ^3\left(\Phi\right) \left[\coth \left(\Phi\right) \left\{4 g^2 \omega_q  \tanh \left(\phi\right) \text{sech}^2\left(\phi\right)+\omega_R^3 \text{csch}^2\left(\Phi\right)\right\}+2 g^2\omega_R \text{sech}^2\left(\phi\right) \text{csch}^2\left(\Phi\right)\right]^2}{\zeta^3 \left[1-\frac{\tanh ^2\left(\Phi\right)}{\zeta^2}\right]},
    \end{aligned}
    \end{equation}
    with
    \begin{equation}
\zeta=4 \eta+\coth \left(\Phi\right).
    \end{equation}
    %------------------------------------Fig 10--------------------------------------------%
The QFI as a function of coupling strength $g$ is plotted in Fig.~\ref{fig10}. We see that the QFI rapidly grows with the increase of $g$ in the weak coupling regime. 
\begin{figure}
    \centering
    \includegraphics[scale=0.6]{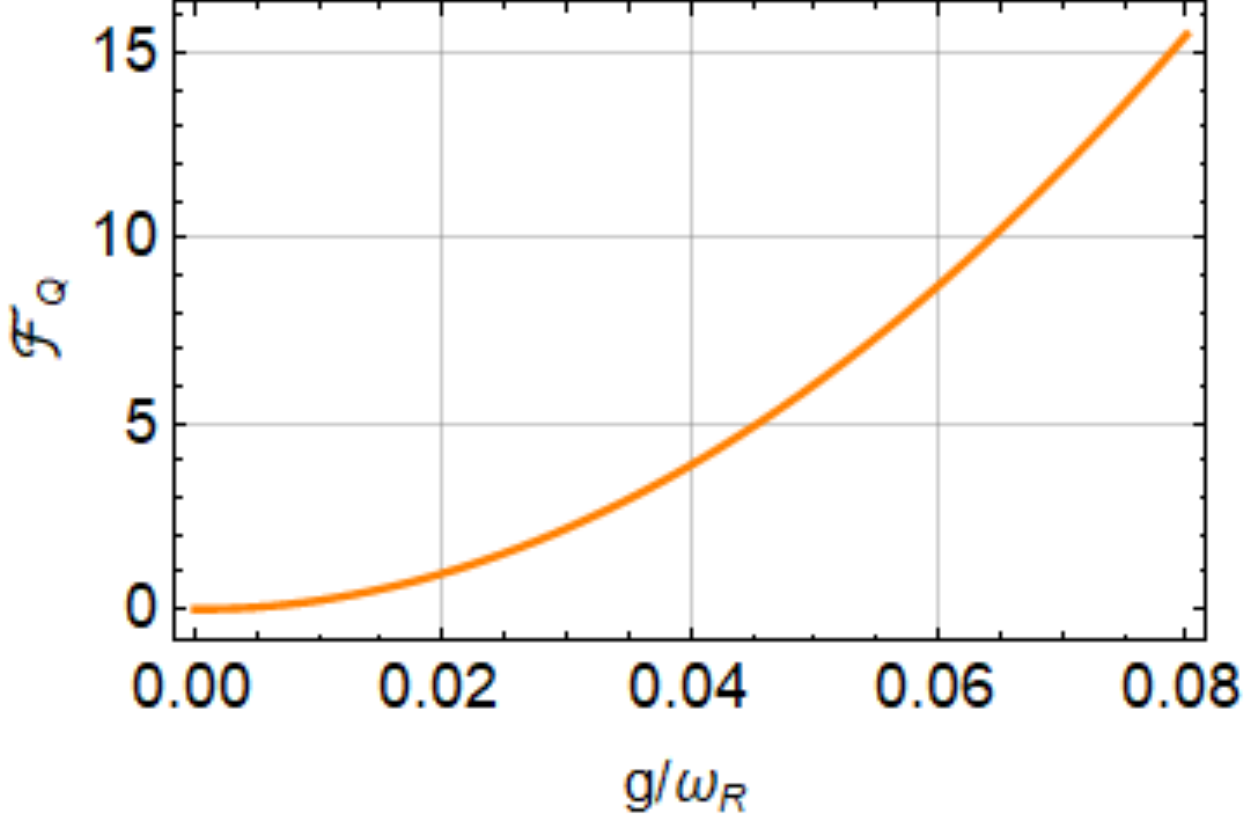}
    \caption{QFI for the probe as a function of coupling strength $g$. The other parameters are set to $\omega_q=0.04$, $\omega_R=1$ and $T=0.02$. The parameters are scaled with the resonator frequency $\omega_R$.}
    \label{fig10}
\end{figure}
    %------------------------------------Fig 10--------------------------------------------%

\end{widetext}
%--------------------------------------------------------------%
\bibliography{CTS.bib}
%--------------------------------------------------------------%
\end{document}